\magnification1200

\rightline{KCL-MTH-04-05}
\rightline{hep-th/0406150}

\vskip .5cm
\centerline
{\bf  $E_{11}$ origin of  Brane charges and U-duality multiplets}
\vskip 1cm
\centerline{Peter West}
\centerline{Department of Mathematics}
\centerline{King's College, London WC2R 2LS, UK}
\vskip 0.5cm
\centerline{and}
\vskip 0.5cm
\centerline{Erwin Schr{\" o}dinger International Institute for
Mathematical Physics,}
\centerline {Boltzmanngasse 9,} 
\centerline{1090 Wien, Austria}

\vskip 2cm
\leftline{\sl Abstract}
\vskip .2cm
\noindent 
We derive general equations which determine the decomposition of the
$G^{+++}$ multiplet of brane charges into the sub-algebras that arise
when  the non-linearly realised $G^{+++}$ theory is dimensionally reduced
on a torus. We apply this to calculate the low level $E_8$ multiplets of
brane charges that arise when the $E_{8}^{+++}$, or $E_{11}$, non-linearly
realised theory is dimensionally reduced to three dimensions on an eight
dimensional torus. We find precise agreement with the U-duality multiplet
of brane charges previously calculated, thus providing a natural eleven
dimensional origin for the "mysterious" brane charges found that do not
occur as central charges in the supersymmetry algebra. We also discuss
the brane charges  in nine dimensions and how they arise
from the IIA and IIB theories. 

\vskip .5cm

\bigskip
{\bf 1 Introduction}
\medskip
Although there are a number of different approaches to string theory it
has been clear for many years that none of them provides a complete
formulation of string theory. For example, string field theory does, at
least in principle, provide a non-perturbative description of string
theory, but it does not  readily accommodate non-trivial backgrounds. The
IIA [1] and IIB [2,3] supergravity theories on the other hand possess so
much supersymmetry that they are essentially unique and as a result must
contain all the low energy effects of the corresponding string theories. 
As such, the properties of these theories have provided much of our
 knowledge of what might constitute a proper formulation of string
theory. The scalars in supergravity multiplets always belong to non-linear
realisations and the groups that occur in these constructions were one of
the most surprising  developments in the construction of   supergravity
theories.    The first time this was observed [4] was in the context of
the $N=4$ supergravity theory in four dimensions,  however, perhaps the
most celebrated example is the
$N=8$ supergravity theory in four dimensions whose scalars belong to the
non-linear realisation of $E_7$ with respect to a  $A_7$ sub-group [5].  
One of the most important such examples for string theory is the SL(2,R)
symmetry [2] of IIB supergravity theory and  one of the  most
interesting from the view point of  this paper concerns  the scalars of
the maximal supergravity theory in three dimensions which belong to a
non-linear realisation of 
$E_8$ with respect to a $D_8$ sub-group [6]. 
\par 
The eleven dimensional
supergravity [7] dimensionally reduced on a circle leads to the IIA
supergravity theory which possess a SO(1,1) coset symmetry, but on a 
$k$   dimensional torus, for
$k\le 8$, it  leads to a theory that possess
$E_{d}$ symmetry   [8].  The IIB supergravity in ten dimensions is not
related to eleven dimensional supergravity in such an obvious way.
However, since there is only one maximally supergravity theory in nine
dimensions, both the IIB supergravity  with the IIA supergravity theories 
in ten dimensions must lead to this  unique maximal theory when  each is
reduced on a circle.  As such,  one need not consider the dimensional
reduction of the IIB theory separately. 
\par
The low energy effective action [9] for the heterotic
string compactified to four dimensions  possess an SL(2,R) symmetry. 
It was realised that this symmetry would be broken by quantum effects 
associated with solitons and it was conjectured that the full quantum
theory would possess an   SL(2,Z)  symmetry.  [10,37] . Furthermore, it
was realised that this symmetry contained what were called S-duality
transformations that swopped perturbative with non-perturbative effects 
and visa versa [10,37].  
It was subsequently proposed [11] that the type II string theories
possessed  the corresponding $E_{d},\ d=1,\ldots, 8$ symmetry found in
the corresponding supergravity theories, but  restricted to be over an
appropriate integer valued field and that the IIB string theory possessed
an SL(2,Z) symmetry. 
These symmetries became known as   U-dualities and
they include the well known T-duality symmetries [12] which are known to
be a symmetry order  by order of   string perturbation theory [13]  and
thought also to hold for  non-perturbative effects. The IIA
string theory reduced on a
$d-1$ dimensional torus is known to be invariant under a T duality
transformation SO(d-1,d-1, Z).  As such, M
theory on a $d$ torus should be  invariant under SO(d-1,d-1, Z). However,
it is also invariant under a SL(d,Z) symmetry which is the remnant of
the general coordinate transformations preserved by the torus. Seen from
the IIB theory, this latter symmetry contains the non-perturbative
S-transformation of SL(2,Z). Hence, M theory reduced on a d torus should
be invariant under the closure of SO(d-1,d-1, Z)  with the SL(d,Z). The 
closure of these two groups  generate  the Weyl group of $ E_{d}$ which
one can take  to define the meaning of $ E_{d}$ taken over the
appropriate field [15,16,17,18].
\par
Carrying out such U-duality symmetry transformations on the solitons in 
the supergravity theories which correspond to the fundamental strings
leads to the solutions for branes and,
as a result, it become clear that  string theory needed extending to a
theory that contains branes as well as strings as fundamental objects. The
branes in eleven dimensional supergravity and the IIA and IIB
supergravity  theories  possess topological charges that occur in the
supersymmetry algebra as central charges [14]. In the  dimensional
reduction of these theories  on a  $d$-dimensional torus the 
branes may    wrap on part of, or sometimes all, of the  torus and one
then finds a more complicated set of branes in the dimensionally reduced
theory. The charges of these branes are given in terms of the one
parameter
$l_{p}$ of the eleven dimensional theory and the radii $R_i,\ d=1,\ldots
,8$ of the torus.  Using the known formulae for the transformations of the
radii and string coupling  under the SO(d-1,d-1) T-dualities and  SL(d,Z)
transformations mentioned above allowed the authors of references
[15,16,17,18]  to  compute the transformations of the brane charges.
Indeed , it was by carry  out this calculation that these authors found 
[15,16,17,18] that the  closure of  these two groups was
isomorphic to the  Weyl group of 
$E_{d}, d=1,\ldots, 8$.  Hence, by construction,  they found a set of
brane charges that belonged to representations  of the Weyl group of
$E_{d}, d=1,\ldots, 8$. 
\par
In view of the result of reference [14] one
might expect the brane charges to occur in the supersymmetry algebra.
Indeed, for a dimensional reduction on
$T^d$ for $d\le 7$ the content of the multiplet of point particle brane
charges  correspond precisely  with the central  charges that occurred in
the supersymmetry algebra of the dimensionally reduced theory [15-18,19]. 
However, for
$d \ge 8$ they found that the  multiplet of brane charges
contained more charges than there were central charges in the
corresponding  supersymmetry algebra. Furthermore, these additional
charges   did not correspond to charges of the familiar branes and
so their meaning and origin was unclear [15,16,17,18,19]. 
For  charges corresponding to strings and
higher dimensional objects the mismatch occurs for dimensional reduction
on a torus of even smaller dimension. 
\par
As explained in  reference [19], these results are not in
conflict with the construction of supergravity theories and their
non-linear realised symmetries. The supercharges  in these  theories
transform under the subgroup, i.e. SO(16) in the case of
$d=8$,  associated with the non-linear realisation and as a result the
central charges must, from the point of view of the supersymmetry
algebra, only transform, and so form multiplets, under this subgroup.
The fact that  for dimensional reduction
on a torus of sufficiently low dimension the central  charges form
multiplets of not just the subgroup, but the 
larger group that occurs in the non-linearly realised
theory, i.e. the Weyl group of $E_{d},
d=1,\ldots, 8$,  suggests that the brane charges should in general form
multiplets of the latter group. Indeed for U-duality to be true this
would have to be the case.  As a result,  it is desirable to find   a
satisfactory origin for the "mysterious" charges that U-duality predicts.
\par
Study of the properties of the  eleven dimensional supergravity theory  
 has lead to the conjecture [20] that M theory possess an $E_{11}$
Kac-Moody symmetry. In particular, it was found that the bosonic sector of
eleven dimensional supergravity theory could be formulated as a non-linear
realisation [21]. The infinite dimensional algebra involved in this
construction was  the closure of a finite dimensional algebra, denoted
$G_{11}$, with the eleven dimensional conformal algebra. The non-linear
realisation was carried out by ensuring that the equations of motion were
invariant under both finite dimensional algebras, taking into
account that some of their generators were in common.  
The algebra
$G_{11}$ involved the space-time translations together with an algebra
$\hat G_{11}$ 
which contained $A_{10}$ and the Borel subalgebra of $E_7$
as subalgebras. The algebra  $\hat G_{11}$  was not a Kac-Moody algebra,
however,  it was conjectured [4] that the theory could be extended so that
the algebra $\hat G_{11}$  was promoted to  a Kac-Moody algebra.  It was
shown that this Kac-Moody symmetry would have to contain a certain rank
eleven Kac-Moody algebra denoted  $E_{11}$ [20]. 
\par
Consequently, it was argued  [20] that an extension of eleven dimensional
supergravity should possess an $E_{11}$ symmetry that was non-linearly
realised. One advantage of this approach is that  the symmetries found
when  the  eleven dimensional  supergravity theory was dimensionally
reduced are already    present in the  eleven dimensional theory and so
occur naturally    . One of the advantages of a   non-linear realisation
is that the dynamics  is largely specified by the algebra if the chosen
local subalgebra is sufficiently large. 
\par
The same analysis was applied to the IIA and IIB supergravity theories
which were conjectured to be part of a larger theory which also possessed
an $E_{11}$ symmetry [20,27]. This is consistent with the idea that the 
type II string theory in ten dimensions and an eleven dimensional theory
are part of a single theory called M theory and indeed their common
$E_{11}$ origin provides  explicit relations between the two theories
[28].   
\par 
Arguments  similar  to those advocated for eleven dimensional
supergravity in [20] were proposed to apply to   gravity [29] in D
dimensions  the effective action of the closed bosonic string [4]
generalised to  D dimensions and the heterotic string [38] and the
underlying Kac-Moody algebras were identified. It was realised that the
algebras that arose in all these theories were of a special kind and were
called very extended Kac-Moody algebras [30]. Indeed,  for any finite 
dimensional semi-simple Lie algebra
$\cal G$ one can systematically extend its Dynkin diagram by adding three
more nodes to obtain an indefinite 
Kac-Moody algebra denoted $\cal G^{+++}$. In this notation $E_{11}$ is
written as $E_8^{+++}$. The algebras for  gravity and the closed bosonic
string being $A^{+++}_{D-3}$ [29] and $D^{+++}_{D-2}$ [20] respectively. 
\par
It was proposed  in [20,29,30,38] and [26,31,32,25], 
  that the non-linear realisation of any  very extended algebra 
${\cal G}^{+++}$ leads to a theory, called ${\cal V}_{\cal G}$ in [32],
that at low levels includes gravity and the other fields and it was 
hoped that this non-linear realisation contains an infinite number of
propagating fields that ensures its consistency. 
Indeed, it was shown [32] that the low level content of the adjoint
representation of $\cal G^{+++}$ predicted a field content for a
non-linear realisation of $\cal G^{+++}$ which was in agreement 
with the oxidized theory associated with algebra $\cal G$. 
\par
Some papers have uncovered
relationships between the solutions in the oxidised theories and the
    $\cal G^{+++}$ symmetry conjectured to be present in their
extension. In reference [26],  the non-linear realisation of $\cal
G^{+++}$ restricted to its Cartan subalgebra was constructed and the
resulting Weyl transformations  were shown to transform the moduli of the
Kasner solutions into each  other. Furthermore,  for
$E^{+++}_8=E_{11}$ and $D_{24}^{+++}$ these Weyl transformations  were
shown to be the   U-duality
transformations in the corresponding string theories. 
\par
The question of how space-time was to be encoded
in the theory was taken up in reference [22] where it was proposed that 
one should take the non-linear realisation of semi-direct product
of $E_{11}$ and a set of generators transforming in the $l_1$
representation of $E_{11}$ where $l_1$ is the fundamental representation
of $E_{11}$  corresponding to the very extended node. The lowest
level generator in this representation is the space-time translation
operator,  at the next two levels it contains the two central charges
that occur in the eleven dimensional supersymmetry algebra and it also
contains an infinite number of higher level object. Hence the first three
objects in the
$l_1$ representation can be interpreted as the  charges
associated with the point particle, two brane and five brane of the eleven
dimensional theory.  It was shown in reference [23] that all the objects
in the
$l_1$ representation have the correct  $A_{10}$ structure to be
interpreted as charges for all the branes  whose sources are fields in
the non-linearly realised theory. An alternative idea for encoding
space was to regard it to arise from the dynamics of $E_{10}$ [24] or 
all of space-time to arise from the dynamics of $E_{11}$ [25].  A
discussion  of the different approaches can be found in reference [23]. 
\par
In this paper we use the approach of reference [22] and take the $l_1$
representation of $E_{11}$ to contain the brane charges. In section two we
derive formulae that can be used to find the low level content of the
$l_1$, and the adjoint representations of $G^{+++}$, in terms of
decompositions of
$G^{+++}$ that are appropriate for the dimensional reduction of the
non-linear realised theory on tori. In section three, we 
calculate the low level content of the 
$l_1$ representation of $E_{11}$ in terms of its $E_8\otimes A_2$
sub-algebra which is the one appropriate to the torus dimensional
reduction  of the non-linearly realised theory to three dimensional
space-time.  As such, we are find at low levels the brane charges of the
non-linearly realised theory. We find in addition to the expected brane
charges which are central charges in the supersymmetry algebra all the
"mysterious charges" found in references [15-18] on the basis of U-duality
considerations. Thus we find that the non-linearly realised $E_{11}$
theory  and its $l_1$ representation provide a natural explanation for
the  additional brane charges that are required if U-duality is to hold
in M theory. 
\par
In section four, we carry out a similar calculation for the decomposition
appropriate to torus dimensional reduction to nine
space-time dimensions and calculate the low level brane charges predicted
by the $E_8^{+++}$ non-linearly theory. We also derive the brane charges
from a IIB perspective and derive relations between these and the  charges
in the IIA theory that  are in agreement at low levels with that
found in references [30,40,19].  We
also find that a truncation of the  $E_8^{+++}$ non-linearly theory in
nine dimensions which makes contact with    the BPS extended theory
considered in references [40,19]. 

%%%%%%%%%%%%%%%%%%%%%%%%%%%%%%%%%%%%%%%%%%%%%%%%%%%%%%%%%%%%%%%%%%%%%%%% 

\bigskip
{\bf 2 Decompositions of  the  $l_1$ representation of $G^{+++}$}
\medskip
As explained in the introduction, the objects
that occur in the $l_1$ representation, that is the fundamental
representation associated with the very extended node, are the  brane
charges for the non-linear realised theory based on $G^{+++}$. In this
section, we derive general equations for calculating the low level content
of the
$l_1$ representation  when decomposed into representations of  the 
sub-algebras of
$G^{+++}$ that arise when the theory is dimensional reduced on a
tori. We will study in detail the most useful
such decomposition  into 
$G\otimes A_2$. This is  one of the most instructive decompositions as it
contains   the largest finite dimensional semi-simple Lie algebra
contained in
$G^{+++}$ and so it allows us to see the most complete structure of  this
representation when expressed  in terms of a Lie algebra that is well
understood. We can think of this as a decomposition of the theory in its
original dimension, but it corresponds to the decomposition that occurs
when the theory is reduced to three dimensional space-time on a eight
torus. In this case, the internal symmetry is $G$ and the 
$A_2$ part is related to the representations of the 
Lorentz algebra in three dimensional space-time.   
\medskip
{\bf 2.1 General decomposition }
\medskip
In  reference [30] is was proposed to study Lorentzian
Kac-Moody algebras which are algebras whose Dynkin diagrams $C$ possess
at least {\bf one }
 preferred node whose deletion leaves a, possible disconnected,  Dynkin
diagram
$C_R$ which is made up of the Dynkin diagrams  all of which are those of
finite dimensional semi-simple Lie algebras
 with possible exception of   at most one affine Lie algebra. The
properties of the Lorentzian Kac-Moody algebra were then studied in terms
of the, possibly reducible, algebra  corresponding to the Dynkin diagrams
that remain after the deletion of the preferred point. 
In this spirit, 
the notion of a level was introduced in the context of
$E_{10}$ in reference [24] and for any Kac-Moody algebra in [33]. The
level of a root of the Kac-Moody algebra is
just and number of times the simple root of the preferred node
enters. The level of a generator being that of its associated root. The
algebra is then studied level by level in terms of the remaining algebra.
The low level generators of
$E_{10}$ and
$E_{11}$  have been studied in terms of $A_{9}$ [24,34] and $A_{10}$
[33,34] respectively by deleting the exceptional node and for all very
extended algebras $G^{+++}$ in [32]. So far  these  level decompositions 
have been studied when the node deleted leaves just one semi-simple finite
dimensional Lie algebra, However, for the application we
wish to consider in this paper, we will delete a node such that the
remaining Dynkin diagram $C_R$ is not irreducible but contains the Dynkin
diagrams of two or more semi-simple finite dimensional algebras $G^{(p)}$.
 The roots, weights and some other quantities  was calculated  in terms
of the remaining algebra in reference [30] for this case, but the
extension to give the  corresponding level decomposition was not given.  
\par
To find the  decomposition of  the $l_1$
fundamental representation of
$G^{+++}$,  
 it is advantageous to consider the adjoint representation of 
$G^{++++}$ [23] from which the former may be extracted in a simple way as
explained below. The latter has a Dynkin diagram that is found from  the
Dynkin diagram
$C$ of 
$G^{+++}$ by adding a new node attached to the very extended node by a
single line. We label the additional node by $*$.
 \par
The non-linear realisation of  $G^{+++}$ contains gravity which is
associated with a preferred $A_{D-1}$ sub-algebra where $D$ is  the  
 space-time dimension in which the resulting theory lives. 
 The Dynkin diagram of this $A_{D-1}$ sub-algebra contains the  very
extended node of the Dynkin diagram of
$G^{+++}$ as well as  
$D-2$ other nodes attached to the very extended node and to each other by
a single line. This line of connected dots has become known as  the
gravity line. For a given $G^{+++}$, there is in general more than one 
possible choice for this sub-algebra, or gravity line. 
\par
In this section
we wish to consider the decomposition of 
$G^{+++}$ and its $l_1$ representation in terms of the sub-algebra that
results from the decomposition of the kth dot along the gravity line
counting from the very extended node. This is the one appropriate to
the decomposition of the non-linear realisation to $k$ space-time
dimensions since the remnant of the gravity line which includes 
the very extended node has 
$k-1$ dots and so corresponds to $A_{k-1}$, or SL(k). In general, the
Dynkin diagram $C_R$ that results from deleting this node  will
contain several pieces which we label by $p,q=1,2,\ldots$, their Dynkin
diagrams being labeled by $C^{(p)}$ and the corresponding sub-algebras
being $G^{(p)}$. The nodes of $C^{(p)}$ are labeled by the indices 
$i,j,\ldots=1,2,\ldots $. The index range will be different for each
$C^{(p)}$ and the ambiguities of labeling of the indices on a given
object are resolved by  the knowledge of Dynkin diagram to
which the object  is associated.  
\par
We label the
simple roots and weights of
$G^{(p)}$ by 
$\alpha_i^{(p)}$ and $\lambda^{(p)}_i$, the range of the index $i$ being
apparent from the object to which it is attached. The Dynkin diagram $C$
of $G^{+++}$ is then labeled in a way which is appropriate to the
decomposition we wish to perform; we label the node which is to be
deleted  by c while the remaining nodes, which belong to  the sub-diagrams
$C^{(p)}$,  are labeled as for the   each sub-diagram. 
In the cases of most interest to us the remaining Dynkin diagram $C_R$
will contain two pieces.  The very extended node will be contained in the 
$C^{(1)}$ diagram. 
\par
As explained above,  we introduce an extra node,  labeled by $*$, 
attached to the very extended node by a single line and consider the
algebra 
$G^{++++}$ so that we may consider the $l_1$ representation of $G^{+++}$.
The roots of  $G^{++++}$  may be written as  
$$
\beta=m_{*}\alpha_{*}+\sum_{a}n_a\alpha_a,
\eqno(2.1.1)$$
where $\alpha_{*}$ is the simple root corresponding to the additional node 
and $\alpha_a$ are the roots of $G^{+++}$. For a positive root $m_*$ and
$n_a$ are positive integers. 
\par
The roots of $G^{++++}$ with 
 level $m_{*}=0$ are just roots of $G^{+++}$. 
Clearly, the commutator of a
level 
$m_{*}=1$ generator with level $m_{*}=0$ generator gives a generator
of level
$m_{*}=1$, as such the $m_{*}=1$ generators form a representation of
$G^{+++}$. It is  the  fundamental representation
with highest weight $\lambda_{1}^{G^{+++}}$ of ${G^{+++}}$ associated with
the very extended node of $G^{+++}$, i.e. the $l_1$ representation. We can
think of
$m_*$ as a level, but we will only be interested in the case $m_*=1$ or
$m_*=0$ .
\par
In order to  carry out the decomposition from $G^{++++}$ to $G^{+++}$ 
we  write 
$$\alpha_{*}=y-\lambda_{1}^{G^{+++}}
\eqno(2.1.2)$$
where  $y$ is a vector that
is orthogonal to the roots of $G^{+++}$. Since 
$\alpha_{*}.\alpha_{*}=2$ and
$\lambda_{1}^{G^{+++}}.\lambda_{1}^{G^{+++}}={1\over 2}$, we conclude
that $y^2={3\over 2}$. 
\par
The decomposition found by deleting the $k$th node of the gravity line
which we referred  to as node c, with simple roots $\alpha_c$, proceeds
 as  explained in reference [30].  We may write the simple roots
of
$G^{+++}$ as 
$$\alpha_c=x-\nu,\ \alpha_i^{(p)},
\eqno(2.1.3)$$
where $\alpha_i^{(p)}$ are the roots of the algebras $G^{(p)}$.  
In the above   
$$\nu=-\sum
_{i,p}A_{ci(p)}^{G^{+++}}\lambda_{i}^{(p)}{(\alpha_c,\alpha_c)\over
(\alpha_{i}^{(p)},\alpha_{i}^{(p)})}
\eqno(2.1.4)$$ where
$A_{ci(p)}^{G^{+++}}$ is the Cartan matrix of $G^{+++}$ labeled as
explained above. This formula differs from that of reference [30] in that
is includes the possibility of the algebra $G$ being non-simply laced. We
note that the simple roots of
$G^{+++}$ for nodes other than c are just the simple roots 
of the sub-algebras
$G^{(p)}$, 
$x$ is orthogonal to all of these and has its length determined by the
requirement that the length of
$\alpha_c$ be as required.  The fundamental weights of
$G^{+++}$ are then given in terms of $x$ and
$\lambda_i^{(p)}$ by 
$$l_i^{(p)}=\lambda_i^{(p)}+{\nu\cdot \lambda_i^{(p)}\over x^2}x,
\ l_c={x\over x^2}
\eqno(2.1.5)$$
We note that $l^{(1)}_1=\lambda_{1}^{G^{+++}}$. 
\par
Using the above expressions, we may write any root $\beta$ of
 $G^{+++}$  as given in equation (2.1.1) in the form  
$$\beta=m_* y+x(m_c-m_* {\nu^{(1)}\cdot \lambda_1^{(1)}\over
x^2})-\sum_p \Lambda^{(p)}
\eqno(2.1.6)$$
where 
$$\Lambda^{(p)}=-\sum_i m_i^{(p)}\alpha_i^{(p)}
+m_c \nu^{(p)}
+\delta _{(1,p)} m_* \lambda_1^{(1)}
\eqno(2.1.7)$$
where $\nu^{(p)}$ is the component of $\nu$ in the sub-algebra
$G^{(p)}$, i.e. $\nu^{(p)}=-\sum
_{i}A_{ci(p)}^{G^{+++}}\lambda_{i}^{(p)}$. 
\par 
We are decomposing the representation of $G^{++++}$
in terms of
$G^1\otimes G^2\otimes \ldots $ and  if a representation of
$G^{(p)}$  occurs in the decomposition of the $l_1$, or adjoint,
representations of $G^{+++}$, then its   
highest weight must occur in  
$\Lambda^{(p)}$ in both  the positive and negative root spaces of
$G^{+++}$. Applying this to the negative roots,  we must find that there
exists a root of $G^{+++}$ such that 
$$ \Lambda^{(p)}=\sum_i p_i^{(p)} \lambda_{i}^{(p)} 
\eqno(2.1.8)$$
where $p_i^{(p)}$ are positive integers. Taking the scalar product with
the fundamental weights of $G^{(p)}$ we then find the conditions 
$$\sum_i p_i^{(p)} \lambda_{i}^{(p)}\cdot  \lambda_{j}^{(p)} 
-m_c \nu^{(p)}\cdot  \lambda_{j}^{(p)}
-\delta _{(1,p)} m_* \lambda_1^{(1)}\cdot  \lambda_{j}^{(1)} 
 =-m_j^{(p)}{2\over
(\alpha_i^{(p)},\alpha_i^{(p)})}
\eqno(2.1.9)$$
where $p_i^{(p)}$, $m_i^{(p)}$, $m_*$ and $m_c$ are all positive
integers. 
The scalar products of the fundamental weights are related to the inverse
Cartan matrices of the sub-algebras by 
$$ (A^{(p)})_{ij}^{-1}={2\over
(\alpha_i^{(p)},\alpha_i^{(p)})}(\lambda_i^{(p)},\lambda_j^{(p)}),\ {\rm
and }\ (\alpha_i^{(p)}, \lambda_j^{(q)} )
={2\delta ^{pq}\delta_{ij}\over (\alpha_i^{(p)},\alpha_i^{(p)})}
\eqno(2.1.10)$$
For finite dimensional semi-simple algebras, the inverse Cartan matrices
are positive definite and as a result the above equation tightly
constrains the possible Dynkin indices $p_i^{(p)} $, or representations
of the sub-algebras $G^{(p)}$ that can arise. 
\par
The above decomposition does not apply when one   deletes the second node
along the gravity line as the fundamental weight associated with this
node has length zero and so can not be written as ${x\over x^2}$. 
\par
The length squared of the roots of $G^{++++}$ which contain the highest
weights of the sub-algebras are given by 
$$\beta^2={3\over 2}m_*^2+x^2(m_c-m_*{\nu^{(1)}\cdot \lambda_1^{(1)}\over
x^2})^2+\sum_q\sum_{i,j} p_i^{(q)}\lambda_{j}^{(q)}
\cdot\lambda_{j}^{(q)}  p_j^{(q)}
\eqno(2.1.11)$$
This quantity is an integer and is bounded from above for all Kac-Moody
algebras and for simply laced Kac-Moody algebras it can only take the
values $2,0,-2,\ldots$. Hence,  we find a further constraint on 
the possible
Dynkin labels
$p_j^{(q)}$ and so representations that can arise. 
\par 
A  hyperbolic Kac-Moody algebras possess roots at
all these values of $\beta^2$, but this is not so for more general
Kac-Moody algebras. For a general Kac-Moody algebra the roots are
either real or imaginary. By definition a  real root not only has
$\beta^2=2$, but must also be conjugate under the Weyl group to  a
simple root. An imaginary root has 
$\beta^2\le 0$, but must also be conjugate under the Weyl group to
 a root that is in the fundamental Weyl chamber and also 
has connected support on the Dynkin diagram of the Kac-Moody algebra.
We consider the integer coefficients of a root when 
expressed  in terms of simple roots.  For those   integer
coefficients that are  zero  we delete 
the corresponding
nodes of the Dynkin diagram. A connected root is one for which 
the resulting  is a connected diagram [36]. We will see that  not every
solution we find  will respect this condition  so we will discard such 
 solutions. 
\par
Although equations (2.1.9) and (2.1.11) are necessary conditions for a
representation to arise they are not sufficient. Indeed, in the case of
the adjoint representation of $G^{+++}$ they do not encode all the
consequences of the Serre relations. However, we know from experience
that the solutions to the above equations usually   arise in the
actual decomposition. We note that the  above formalism does not predict
the number of times a representation can arise at a given level. 
\par
The most common case, and the one of most interest to us, is when the
deletion of the node on the gravity line leads to only two sub-algebras. 
It is useful in this case to introduce a more easily understood notation. 
We denote the roots,  weights and Dynkin indices of the  sub-algebras by 
$$\alpha_i^{(1)}=\beta_i ,\ \lambda_i^{(1)}=\mu_i,\ p_i^{(1)}=q_i,\
m_i^{(1)}=m_i;\ \ 
 \alpha_i^{(2)}=\alpha_i ,\ \lambda_i^{(2)}=\lambda_i,\ p_i^{(2)}=p_i
,\ m_i^{(2)}=n_i. 
\eqno(2.1.12)$$ 
We will also restrict our attention from now on to when $G$ is simply
laced and so all the simple roots of $G^{+++}$ have length squared two. In
this notation equation (2.1.9) becomes 
$$\sum_i q_i \mu_{i}\cdot  \mu_{j}
-m_c \nu^{(1)}\cdot  \mu_{j}
- m_* \mu_1\cdot  \mu_{j}=-m_j
\eqno(2.1.13)$$
for the first sub-algebra and 
$$\sum_i p_i \lambda_{i}\cdot  \lambda_{j} 
-m_c \nu^{(2)}\cdot  \lambda_{j}=-n_j
\eqno(2.1.14)$$
for the second sub-algebra. 
%%%%%%%%%%%%%%%%%%%%%%%%%%%%%%%%%%%%%%%%%%%%%%%%%%%%%%%%%%%%%%%%%%%%%%%%%%
\medskip
{\bf 2.2 Decomposition  into  $G\otimes A_2$}
\medskip
In this
section, we derive the equations that allow the calculation of  the
content of the
$l_1$ representation  decomposed into representations of $G\otimes A_2$.
This is the decomposition into the largest finite dimensional semi-simple
Lie algebra contained in $G^{+++}$ and so it allows us to see the most
complete structure of  this representation when expressed  in terms of a
Lie algebra that is well understood. We can think of this as a
decomposition of the theory in its original dimension, but it corresponds
to the decomposition that occurs when the theory is dimensionally reduced
 to three dimensional space-time. In this case, the
internal symmetry is $G$ and the 
$A_2$ part is related to the representations of Lorentz algebra
in three space-time dimensions.  As explained in appendix B, the Lorentz
group arises  essentially as the Cartan involution invariant sub-algebra
of $A_2$. The $A_2$ representations lead to representations of the
Lorentz algebra that transform under the Lorentz algebra as the $A_2$
indices might naively suggest.  
\par
This decomposition results from the  deletion of  the third node in the
gravity line. The gravity line begins at the very extended node of 
 $G^{+++}$ and then
contains the over extended node and then the affine node which is the
one to be deleted. Since by construction no other nodes are attached 
to the very extended and over extended nodes in
the Dynkin diagram of $G^{+++}$  this deletion leads to the sub-algebra
$G\otimes A_2$ as claimed. As explained
above,  we label the affine node by c, the  very extended and over
extended nodes of 
$G^{+++}$ which make up the algebra $A_2$ by
$1,2$ respectively and label the nodes of  $G^{+++}$ which form $G$ after
the deletion by
$1,2,\ldots$. 
\par
Following equation (2.1.3), we take   
$$
\alpha_c=x-\nu,\ {\rm where }\ 
\nu=\nu^{(1)}+\nu^{(2)},\ {\rm and}\ \nu^{(1)}=\mu_2, \ 
\nu^{(2)}=\theta=\sum_j c_j\lambda_j
\eqno(2.2.1)$$
In this equation $\theta $ is the highest root of $G$ and we have used
the relation that $A^{G}_{cj}=-(\theta, \alpha_j)$ as c is the affine
node of $G$. For the simply laced algebras we are studying here
$\theta^2=2$. Since $\alpha_c^2=2$ we find that $x^2=-{2\over 3}$. 
\par
Equation (2.1.14) can be written as 
$$\sum_ip_i ((A^G)^{-1})_{ij}-m_c\sum_k c_k ((A^G)^{-1})_{kj}=-n_j
\eqno(2.2.2)$$
for  $p_i, n_j=0,1,2,\dots$ for fixed value of the level
$,m_c=0,1,2,\ldots$.  While equation (2.1.13) becomes 
$$ \sum_{i=1}^2 q_i
((A^{A_2})^{-1})_{ij}- m_c((A^{A_2})^{-1})_{2j}-
m_*((A^{A_2})^{-1})_{1j}=- m_j
\eqno(2.2.3)$$
for  $q_i, m_j=0,1,2,\ldots$ 
for fixed value of the level
$m_{*},m_c=0,1,2,\ldots$. 
As explained above,  since the inverse Cartan matrix of any finite
dimensional semi-simple Lie algebra is positive definite equation
(2.2.2) allows only a finite number of solutions for a given $m_c$. We
note that   it does not  depend on
$m_*$.  As such, we may calculate, at low
levels, the possible $G$ representations 
 present using this equation. 
\par
Explicitly writing out equation
(2.2.3) we find that it becomes
$$
2q_1+q_2-2m_{*}-m_c=-3m_{1},\ q_1+2q_2-m_{*}-2m_c=-3m_{2}
\eqno(2.2.4)$$
As the $m_i$ and $q_i$ are positive, and  as
previously observed, there clearly are only a finite number of solutions 
for fixed $m_c$ and $m_{*}$.
\par
Furthermore, as $\beta$ is a root of $G^{++++}$ it must have length
squared $2,0,-2,\ldots$  and so equation (2.1.11) becomes  
$$\beta^2=a+\sum_{i,j}p_i((A^G)^{-1})_{ij}p_j
+\sum_{i,j=1}^2 q_i((A^{A_2})^{-1})_{ij}q_j=2,0,-2,\ldots
\eqno(2.2.5)$$
where $a={3\over 2}m_{*}^2-{1\over 6}(m_{*}+2m_{c})^2$. 
 \par
We are interested in the decomposition of the
$\l_{1}$ representation of ${G^{+++}}$  for which we take $m_{*}=1$. 
For $m_{*}=1$ and $m_c=1,2,3,4$  the solutions to equation (2.2.3), or
equivalently (2.2.4),  are  given in table 2.1

$$\vbox{ \offinterlineskip\halign{
\strut#&\vrule#\quad& 
\hfil$#$\hfil&
\quad\vrule#\quad&
\hfil$#$\hfil&
\quad\vrule#\quad&
\it#\hfil&\quad\vrule#\quad&
\hfil#&\quad\vrule#\cr
\noalign{\hrule} 
& &\multispan5\hfil 2.1 Solutions of the $A_2$ equation for $m_{*}=1$
\hfil & & \hfil 
& \cr
\noalign{\hrule} 
& & \omit\hidewidth $m_c$ \hidewidth& & \omit\hidewidth
$(q_1,q_2)$\hidewidth& &
\omit\hidewidth $(m_{1},m_{2})$\hidewidth 
& & \omit\hidewidth $\sum qA^{-1}q$ \hidewidth & \cr 
\noalign{\hrule}
& & 0&  &(1,0) & &(0,0) &&${ 2\over 3}$ &\cr \noalign{\hrule}
& & 1& & (1,1) & &(0,0) && $2$  &\cr \noalign{\hrule}
& & 1&  &(0,0) & &(1,1) &&$0$  &\cr \noalign{\hrule}
& & 2&  &(0,1) & &(1,1) && ${ 2\over 3}$  &\cr \noalign{\hrule}
& & 2&  &(1,2) & &(0,0) && $4+{ 2\over 3}$  &\cr \noalign{\hrule}
& & 2&  &(2,0) & &(0,1) && $2+{ 2\over 3}$  &\cr \noalign{\hrule}
& & 3&  &(1,0) & &(1,2) && ${ 2\over 3}$  &\cr \noalign{\hrule}
& & 3&  &(0,2) & &(1,1) && $2+{ 2\over 3}$  &\cr \noalign{\hrule}
& & 3&  &(2,1) & &(0,1) && $4+{ 2\over 3}$  &\cr \noalign{\hrule}
& & 3&  &(1,3) & &(0,0) && $8+{ 2\over 3}$  &\cr \noalign{\hrule}
& & 4&  &(0,0) & &(2,3) && $0$  &\cr \noalign{\hrule}
& & 4&  &(0,3) & &(1,1) && $6$  &\cr \noalign{\hrule} 
& & 4&  &(1,1) & &(2,1) && $2$  &\cr \noalign{\hrule}
& & 4&  &(1,4) & &(0,0) && $6+{2\over 3}$  &\cr \noalign{\hrule}
& & 4&  &(2,2) & &(1,0) && $8$  &\cr \noalign{\hrule}
}} 
$$
\par
We also give the solutions at low levels to $A_2$ equation (2.2.4)
for $m_*=0$ which corresponds to the adjoint representation of $G^{+++}$. 
There are in fact no solutions for $m_c=0,1$, the higher level solutions
are given in the table 2.2  below

$$\vbox{ \offinterlineskip\halign{
\strut#&\vrule#\quad& 
\hfil$#$\hfil&
\quad\vrule#\quad&
\hfil$#$\hfil&
\quad\vrule#\quad&
\it#\hfil&\quad\vrule#\quad&
\hfil#&\quad\vrule#\cr
\noalign{\hrule} 
& &\multispan5\hfil 2.2 Solutions of the $A_2$ equation for $m_{*}=0$
\hfil & & \hfil 
& \cr
\noalign{\hrule} 
& & \omit\hidewidth $m_c$ \hidewidth& & \omit\hidewidth
$(q_1,q_2)$\hidewidth& &
\omit\hidewidth $(m_{1},m_{2})$\hidewidth 
& & \omit\hidewidth $\sum qA^{-1}q$ \hidewidth & \cr 
\noalign{\hrule}
& & 2&  &(0,2) & &(0,0) &&$2+{ 2\over 3}$ &\cr \noalign{\hrule}
& & 2& & (1,0) & &(0,1) && ${ 2\over 3}$  &\cr \noalign{\hrule}
& & 3&  &(0,3) & &(0,0) &&$6$  &\cr \noalign{\hrule}
& & 3&  &(1,1) & &(0,1) && ${ 2}$  &\cr \noalign{\hrule}
& & 3&  &(0,0) & &(1,0) && $0$  &\cr \noalign{\hrule}
}} 
$$
\par
Equation (2.2.2) is much more complicated to solve. However, the results
only depend on the  level $m_c$, the inverse Cartan matrix of $G$ and the
highest root $\theta$, or equivalently the coefficients $c_k$, of
$G$. For simply laced algebras, the highest root $\theta=\nu ^{(2)}$ can
be expressed in terms of the fundamental weights as follows  
$$\vbox{\tt \halign{\hfil #\hfil && \quad \hfil #\hfil \cr
$G$     &$ E_8$    &$E_7$ &$E_6$ &$A_n$& $D_n$ &$D_3$\cr
$\theta$& $\lambda_1$& $\lambda_6$& $\lambda_6$& $\lambda_1+\lambda_n$& 
$\lambda_2$ & $\lambda_2+\lambda_3$\cr}}
$$
The only solution for level
$m_{c}=0$ is $p_i=0$, while for $m_{c}=1$ there is the obvious solution
namely 
$p_k=c_k$. For both of these solutions  $n_k=0$. 
More solutions for the case of $E_8$ are discussed at the end of this
section. 
\par
Having solved equations (2.2.2) and (2.2.4) we can substitute the results
for
$q_i$ and $p_i$ into the remaining constraint of equation (2.2.5) and see
which of these solutions are allowed. The values of the first term $a$ in
this equation for    the
$l_1$ representation, i.e. $ m_{*}=1$,  are  listed  below for low values
of
$m_c$: 
$$
\vbox{\tt \halign{\hfil #\hfil && \quad \hfil #\hfil \cr
$m_c$     &0    &$1$ &$2$ &$3$& $4$ \cr
$a$& $1+{1\over 3}$& $0$& $-2-{2\over 3}$&
$-6-{2\over 3}$& 
-12 \cr}}
$$
\par
Taking $m_{c}=0$, we find that equation (2.2.4) can only be solved for
$(q_1=1,q_2=0)$,  while the only solution to equation (2.2.2) is $p_i=0$.
We find that these do indeed solve equation (2.2.5) for a $G^{++++}$ root
which is given by   
$\beta=(1,0, \ldots ,0)$ which has $\beta^2=2$. This corresponds to
an element  which is a one rank tensor  of $A_2$ and so a vector of the
space-time group SO(1,2), but is inert under $G$. This root is the
highest weight
$A_{10}$ state of $G^{++++}$ corresponding to $P_1$. As a result, 
the solution we have found 
can be identified with
$P_a,
a=1,2,3$. 
\par
Taking $m_{c}=1$, we have the solution $p_k=c_k$ for equation
(2.2.2) for which
$\Lambda=\theta$ and so $\sum pA^{-1}p=2$. As a result, only one of the
two solutions to equation (2.2.4),  listed in table 2.1, is allowed in
equation (2.2.5) namely for the values 
$(q_1=0,q_2=0)$. The corresponding  $G^{+++}$  root is given
by  
$\beta=(1,1,1,1,c_1,c_2,\dots)$ which has $\beta^2=2$. This is a scalar
under SO(1,2), but has highest weight $\theta$ of  $G$ and  so belongs to 
the adjoint representation of $G$. 
\par
For future use we give in the table 2.3 below the solutions to equation
(2.2.4) for the case of $E_{8}$  up to level four. 

$$\vbox{ \offinterlineskip\halign{
\strut#&\vrule#\quad& 
\hfil$#$\hfil&
\quad\vrule#\quad&
\hfil$#$\hfil&
\quad\vrule#\quad&
\it#\hfil&\quad\vrule#
%\quad&\hfil#&\quad
\vrule\cr
\noalign{\hrule} 
& &\multispan5\hfil 2.3 Solutions of the $E_8$ equation 
\hfil&\cr
\noalign{\hrule} 
& & \omit\hidewidth $m_c$ \hidewidth & &
\omit\hidewidth $p_i$\hidewidth 
& & \omit\hidewidth $\sum pA^{-1}p$ \hidewidth & \cr 
\noalign{\hrule}
& & 0& &p_1=0 && $0$  &\cr \noalign{\hrule}
& & 1 & &p_1=1 &&$2$  &\cr \noalign{\hrule}
& & 1 & &p_i=0 && $0$  &\cr \noalign{\hrule}
& & 2 & &p_1=1 && $2$  &\cr \noalign{\hrule}
& & 2 & &p_1=0 && $0$  &\cr \noalign{\hrule}
& & 2 & &p_7=1 && $4$  &\cr \noalign{\hrule}
& & 3 & &p_1=1 && $2$  &\cr \noalign{\hrule}
& & 3 & &p_7=1 && $4$  &\cr \noalign{\hrule}
& & 3 & &p_8=1 && $8$  &\cr \noalign{\hrule}
& & 3 & &p_2=1 && $6$  &\cr \noalign{\hrule}
& & 3 & &p_i=0 && $0$  &\cr \noalign{\hrule}
& & 4 & &p_2=2 && $24$  &\cr \noalign{\hrule}
& & 4 & &p_3=1 && $12$  &\cr \noalign{\hrule}
& & 4 & &p_4=1 && $20$  &\cr \noalign{\hrule}
& & 4 & &p_6=1 && $14$  &\cr \noalign{\hrule}
& & 4 & &p_7=1 && $4$  &\cr\noalign{\hrule} 
& & 4 & &p_7=2 && $16$  &\cr \noalign{\hrule}
& & 4 & &p_8=1 && $8$  &\cr\noalign{\hrule} 
& & 4 & &p_2=1=p_7 && $18$  &\cr \noalign{\hrule}
& & 4 & &p_2=1 && $6$  &\cr\noalign{\hrule} 
& & 4 & &p_i=0 && $0$  &\cr\noalign{\hrule}
& & 4 & &p_1=1=p_2 && $14$  &\cr\noalign{\hrule}
& & 4 & &p_1=1=p_3 && $22$  &\cr\noalign{\hrule} 
& & 4 & &p_1=1=p_7 && $10$  &\cr\noalign{\hrule}
& & 4 & &p_1=1=p_8 && $16$  &\cr\noalign{\hrule}
& & 4 & &p_1=1 && $2$  &\cr\noalign{\hrule}
}} $$
\par
We also have the solutions  
$p_1=m_c-r$, all other $p_i$'s are zero, with $\sum pA^{-1}p=2(m_c-r)^2$, 
whenever $p_1$ is positive and for integer $r$. For $p_1-m_c=1$ these are
the only other solutions, but  for $p_1-m_c=2$ we can also have 
 the solution with $p_2=1$,  for which 
$\sum pA^{-1}p=2(m_c-2)^2 +6(m_c-2)+6$, and the solution $p_7=1$ for
which 
$\sum pA^{-1}p=2(m_c-2)^2 +4(m_c-2)+4$. In fact, we have
included some of these solutions above where it was useful to do so. 

%%%%%%%%%%%%%%%%%%%%%%%%%%%%%%%%%%%%%%%%%%%%%%%%%%%%%%%%%%%%%%%%%%%%%%
\medskip
{\bf 3 Brane charges in three dimensions}
\medskip
It has been known for many years that eleven dimensional supergravity
dimensionally reduced on a eight-dimensional  torus leads to a theory in
three space-time dimensions that possess an $E_8$ symmetry. [6]. It has
been conjectured that there should exist an extension of 
the maximal supergravity theory in three space-time dimensions that
includes string as well as branes that is invariant under an $E_8$
symmetry  defined over an appropriate integer field [11], called
U-duality. 
\par
It has been proposed [15-18] that the U-duality transformations should be 
generated by the SL(8,{\bf Z}) remnant of general coordinate
transformations preserved by the eight torus with radii $R_a, a=1,\ldots
8$; 
$$ R_a \leftrightarrow R_{a+1},\  a=1,\ldots, 7, 
\eqno(3.1)$$
together with the double 
T-duality transformations of the type IIA theory found after the
reduction on the first circle
$$ R_a\to {l^2_s\over R_a},\  R_b\to {l^2_s\over R_b},\ 
 g_s\to {g_s l^2_s\over R_a R_b}, \ a,b=2,\ldots ,8
\eqno(3.2)$$
all other radii unchanged. Here $l_s$ is the string scale and $g_s$ is
the string coupling constant.  Using the relations $l_p^3=g_s l^3_s$  and
$R_1=g_s l_s$ which relate
the eleven dimensional Planck length $l_p$ and the radius $R_1$ of the
circle used to reduce to the IIA theory we can relate the IIA variables
to those of eleven dimensions to rewrite equation (3.2) as 
$$ R_a\to {l^3_p\over R_b R_c},\  R_b\to {l^3_p\over R_c R_a},\ 
 R_c\to {l^3_p\over R_a R_b},\  l_p^3\to {l^6_p\over R_a R_b R_c},\ 
, \ a,b,c=1,\ldots ,8
\eqno(3.3)$$
all other radii being unchanged. In deriving this last equation we have 
used the possibility to swop radii using equation (3.1). 
It has been shown [15-18] that the closure of the transformations of
equations (3.1) and (3.3) is a group that is isomorphic to the Weyl
transformations of $E_8$ and observed that in this approach one defines a
symmetry which is automatically over an integer field. 
\par
The dimensional reduction of the point particle, two and fives branes of
the eleven dimensional theory lead to  a more rich structure of branes
in three dimensions as these branes may wrap around 
 different directions of the eight-dimensional torus in the dimensional
reduction procedure. The charges  of the
branes in the three dimensional theory should  belong to multiplets of the
Weyl group of
$E_8$, as well as belong to representations  of  SO(1,2) Lorentz
symmetry of space-time, if the conjectures on U-duality are to be true.
Starting with the known charges that do arise from dimensional reduction
of the branes of the eleven dimensional supergravity theory,  the authors
of reference [15-18] used the U-duality transformations of equation (3.1)
and (3.2)   to find   the complete $E_8$ Weyl group charge multiplets 
for the point particles and strings of the three dimensional theory. They
found that the point particle charges belong to the adjoint, or
$\lambda_1$, representation of $E_8$ while the string charges belong to
the  3875, or $\lambda_7$ representation, of $E_8$. 
However, the authors of reference [15-18] found that these $E_8$ Weyl
multiplets of brane charges contained more brane charges that one would
expect from the dimensional reduction of the branes of  the eleven
dimensional supergravity theory.   Put another way the brane charges that
arise from the dimensional reduction  did not form
multiplets of
$E_8$, and so  some of the brane charges in the
$E_8$ Weyl group multiplet did not have an origin in the eleven
dimensional supergravity theory. In particular, the brane charges have
index structures that do not correspond to the coupling to any of the
fields in the eleven dimensional supergravity theory. 
\par
The point particle, two and five branes of the eleven dimensional 
supergravity theory have charges that occur as central charges in 
the eleven dimensional supersymmetry algebra and indeed all such central 
charges have this interpretation. The dimensional reduction of these
branes lead to branes whose charges also occur as central charges in the
dimensional reduced supersymmetry algebra. Hence, the discovery mentioned
above implies that the central charges of the dimensionally reduced
supersymmetry algebra do not belong to $E_8$ Weyl multiplets. 
 Although the supergravity theory is $E_8$ invariant, the 
gravitino, and so the spinorial supercharges transform, linearly only 
under the sub-algebra SO(16). Hence, from the
perspective of the supergravity theory, there is no requirement for the
central charges to belong to multiplets of $E_8$. 
\par
The above consideration apply just as well to dimensional reductions  
on tori of less than eight dimensions where the corresponding algebra
is $E_d, d\le 8$. 
\par 
In this paper we adopt the viewpoint advocated in
[20], namely that M theory is a non-linear realisation of $E_8^{+++}$. 
The fields of M theory belong to the adjoint representation of
$E_8^{+++}$. At low levels these are precisely the fields of eleven
dimension supergravity, including their duals, [20] and it is hoped that
the fields at higher levels are dynamical and ensure the consistency of
the theory. It was shown [26] that the Weyl transformations $E_8^{+++}$
lead to transformations of the fields in the non-linear realisation that
lead precisely to the transformations of equations (3.1) and (3.2). This
is of course consistent with the work of references [15-18], but [26] 
provides a  derivation of these formulae based on the
non-linearly realised $E_8^{+++}$ in contrast to the  origins of these
formulae in reference [15-18] which used the stringy  property such as 
T-duality.  In fact, the transformations of equation (3.1) are Weyl
transformations corresponding to the simple roots on the gravity line and
the transformations of equation (3.3) is just the Weyl transformation of
the simple root of the exceptional node.  
\par
 The
fundamental representation of $E_8^{+++}$ associated with its very
extended node 
$l_1$, contains the space-time generators $P_{\hat a}, \hat a=1,\ldots
,11$, which are the charges for the point particle of M theory,  and the
next two components in the
$l_1$ multiplet  contain the central charges of the eleven dimensional
supersymmetry algebra which are the charges of the two and five brane 
respectively [22]. As such, we may conclude that this multiplet contains
the "brane" charges of the full non-linearly realised 
$E_8^{+++}$ theory. Indeed, as discussed in
[23],  it contains the correct  $A_{10}$ representations to be identified
with the  charges of all the  solutions to the non-linear theory including
those which  are beyond those found in the eleven dimensional supergravity
approximation. The decomposition of the
$l_1$  representation into $E_8\otimes A_2$ representations was
studied in the previous section. It   is the one appropriate for the
dimensional reduction of the theory on a eight torus to three
space-time dimensions and it allows us to read off the brane charges of
the three dimensional theory in terms of multiplet of
$E_8\otimes A_2$.  As discussed in appendix B, the  latter factor, $ A_2$,
is  related to the three dimensional Lorentz algebra. Since the $l_1$
representation has an infinite number of states, it will contain an
infinite number of 
$E_8\otimes A_2$ representations which are classified according to the
level $m_c$. It is of course inevitable  in this approach that the brane
charges in three dimensions belong to multiplets of $E_8$. 
\par 
While the previous section provides a systematic analysis of the
decomposition we require, it is also instructive to consider the $l_1$ 
representations as it occurs in eleven dimensions and graded
according to the level $n_8$, often called $n_{11}$ in previous works,
which is taken to be the number of time the root $\alpha_8$ occurs. The
states are then  classified according to the representations of the
algebra that remains once the node $8$ is deleted, namely 
$A_{10}$. This was carried out for low levels in  
 reference [22,23] and, for convenience, we
recall these low level  charges  
$$P_{\hat a} \ (0,2),\  Z^{\hat a_1\hat a_2}\ (1,2),
\  Z^{\hat a_1\ldots \hat a_5}\ (2,2)
,\  Z^{\hat a_1\ldots \hat a_7,b}\ (3,2),
\  Z^{\hat a_1\ldots \hat a_8}\ (3,0),
$$
$$
\  Z^{\hat b_1\hat b_2 \hat b_3, \hat a_1\ldots \hat a_8}\ (4,2),
\  Z^{(\hat c \hat d ), \hat a_1\ldots \hat a_9}\ (4,2),
\  Z^{\hat c\hat d,\hat a_1\ldots \hat a_9}\ (4,0),\ 
\  Z^{\hat c,\hat a_1\ldots \hat a_{10}}\ (4,-2),\ 
Z\ (4,-4)
$$
$$
Z^{\hat c, \hat d_1\ldots \hat d_4,\hat a_1\ldots \hat a_9}\ (5,2),\ 
Z^{\hat c_1\ldots \hat c_6,\hat a_1\ldots \hat a_8}\ (5,2),\ 
Z^{\hat c_1\ldots \hat c_5,\hat a_1\ldots \hat a_9}\ (5,0),\ 
$$
$$
Z^{\hat d_1,\hat c_1 \hat c_2 \hat c_3,\hat a_1\ldots \hat a_{10}}\
(5,0),\  Z^{\hat c_1 \ldots \hat c_4,\hat a_1\ldots \hat a_{10}}\
(5,-2),\  Z^{(\hat c_1\hat c_2,\hat c_3 )}\ (5,2),\ 
Z^{\hat c,\hat a_1\hat a_2}\  (5,-2),\ 
$$
$$ 
Z^{\hat c_1\ldots \hat c_{3}}\ (5,-4),\ 
\eqno(3.4)$$
In the above, the first figure in the brackets refers to the level
$n_8$, while  the second figure is the length squared $\beta^2$ of the
root in
$E_8^{++++}$ to which the highest weight of $A_{10}$ representation
belongs. The index range is $\hat a,\hat b, \ldots =1,\ldots , 11$. The
actual roots
$\beta$ can be found in reference [23]. We note that lower down in the
list we find the state
$$ 
Z^{\hat d,\hat c_1 \dots \hat c_8,\hat a_1\ldots \hat a_{8}}\ (6,2)
\eqno(3.5)$$
\par
Given the above list we may  carry out the dimensional reduction
to three dimensions "by hand" by dividing the index range $\hat a=(a,i)$
where 
$i,j,\ldots =1,\ldots 8$ and
$a,b,\ldots =1,2,3$. The first set are $A_7$, or SL(8) indices, while the
latter are  the SL(3) indices. Although the eleven dimensional origin of
the resulting states is clear from this method, the way the states package
up  into representations of $E_8$ is less clear. However, the above
discussion will help us identify the eleven dimensional origin of the
brane charges that we find in three space-time dimensions. 
\par 
We now return to the $E_8\otimes A_2$ decomposition of the $l_1$
representation in terms of the level $m_c$.  For $m_c=0$, we only
have the solution discussed in the previous section, namely the only
solution has $p_i=0$ and $(q_1,q_2)=(1,0)$ with an $E_8^{++++}$ root of 
$\beta=(1,0,\ldots, 0)$  which has $\beta^2=2$. This is a singlet under
$E_8$ and a vector under the three-dimensional Lorentz algebra; it is
just the
$P_a, \ a=1,2,3$ in the first line of equation (3.4).  

\par
The solutions for $m_c=1$ to equations (2.2.2), (2.2.3) and (2.2.5) for
the case of $G=E_8$ are listed in table 3.1 given below. 

$$\vbox{ \offinterlineskip\halign{
\strut#&\vrule#\quad& 
\quad\quad\hfil#\hfil&
\quad\vrule#\quad&
\hfil#\hfil&
\quad\vrule#\quad&
\it#\hfil&\quad\vrule#\quad&
\hfil#&\quad\vrule#\cr
\noalign{\hrule} 
& &\multispan4\hfil 3.1 Solutions for $m_c=1$  \ \ 
\hfil & & \hfil &&
\cr
\noalign{\hrule} 
& & \omit\hidewidth $E_8\otimes A_2$ \hidewidth & & \omit\hidewidth
$\beta$\hidewidth & &
\omit\hidewidth $\beta^2$\hidewidth 
& & \omit\hidewidth charge\hidewidth & \cr 
\noalign{\hrule}
& & $\lambda_1\otimes (0,0) $  &  &$\beta=(1,1,1,1,0,0,0,0,0,0,0,0)$ &
&2 & &$Z^{i_1\ldots i_7}$ &\cr
\noalign{\hrule} 
& & $1\otimes (0,0)$& & $\beta=(1,1,1,1,2,3,4,5,6,4,2,3)$ &
&0 &&
$Z$  &\cr
\noalign{\hrule} 
& & $1\otimes (1,1)$&  &$\beta=(1,0,0,1,2,3,4,5,6,4,2,3)$ & &2
&&$Z^a{}_b$  &\cr \noalign{\hrule} 
}} 
$$

The first column shows the $E_8\otimes A_2$ representation content, the
second column the $E_8^{++++}$ root $\beta$ for the highest weight state
of 
$E_8\otimes A_2$, the third column gives $\beta^2$ and the final column
displays the the Sl(8) and SO(1,2) indices of the highest weight state. 
\par
The first multiplet  in  table 3.1 is a SO(1,2) singlet and so corresponds
to charges of point particles in the three dimensional theory. They belong
to   the 248  representation of
$E_8$ which decomposes under SL(8) as 
$248\to 8+28+56+(63+1)+56+28+8$. By examining the list of eleven
dimensional charges of equation (3.4) and their $E_8^{++++}$ roots given
in reference [23]  we can find how the point particle charges of the three
dimensional arise in the eleven dimensional $E_8^{+++}$ non-linearly
realised theory. The highest weight component in table 3.1   has an
$E_8^{++++}$  root of $(1,1,1,1,0,0,0,0,0,0,0,0)$.  and we see that this
must be identified with the state $P_4$ in equation (3.4) which arises
from the
$A_{10}$  highest weight state $P_1$ with root $(1,0^{11})$ by the action
of $K^1{}_2, K^2{}_3$ and $ K^3{}_4$. Hence, the first $8$ of the
$248$ arise from the eleven dimensional theory as $P_a,\ a=4,\ldots ,11$.
Similar considerations  allow us  to identify the 248 states and their
$E_8^{++++}$ roots in terms of the eleven dimensional brane charges of
equation (3.4). Looking at the listing of roots on page 22 of reference
[23] we must look for roots that have $(1^4,n_1,n_2,\ldots ,n_8)$ and are
graded according to the construction of the root string of the adjoint
representation of $E_8$. The relevant roots are easy to spot and are as
follows; 
$$P_i\ (8) (1^4,0^8),\ Z^{ij} (28) (1^9,0^2,1),\ 
Z^{i_1\ldots i_5} (56) (1^7,2,3,2,1,2),\ 
$$
$$
Z^{i_1\ldots i_7,j} (63) (1^5,2,3,4,5,3,1,3),\ 
Z^{i_1\ldots i_8} (1) (1^4,2,3,4,5,6,4,2,3),\ 
$$
$$
Z^{i_1\ldots i_8,j_1\ldots j_3} (56) (1^4,2,3,4,5,6,4,2,4),\ 
$$
$$Z^{i_1\ldots i_6,j_1\ldots j_8} (28) (1^4,2,3,5,7,9,6,3,5),\ 
Z^{i_1\ldots i_8,j_1\ldots j_8,k} (8) (1^4,3,5,7,9,11,7,3,6),\ 
\eqno(3.6)$$
We note that the final $8$ of the 248 occurs at a level of $n_8=6$ 
and so it is far above the branes  whose charges occur in the 
eleven dimensional    supersymmetry
algebra.   Indeed, the corresponding branes   will couple 
to fields of the non-linear realisation  which are well  beyond those
found in the usual supergravity approximation; for example,  the branes
associated with the final 
$8$ couple to a field at level six  in the non-linearly realised theory. 
\par
The second multiplet  in table 3.1 is a singlet under both $E_8$ and
SO(1,2). The final multiplet listed in table (3.1) has an
$E_8^{++++}$  root which does not have connected support on the
$E_8^{++++}$ Dynkin diagram and, as previously explained, it is  not
an acceptable root and so may be discarded. 
\par
The solutions for $m_c=2$ 
to equations (2.2.2), (2.2.3) and (2.2.5) for
the case of $G=E_8$ are listed in table 3.2 given below. 

$$
\vbox{ \offinterlineskip\halign{
\strut#&\vrule#\quad& 
\quad\quad\hfil#\hfil&
\quad\vrule#\quad&
\hfil#\hfil&
\quad\vrule#\quad&
\it#\hfil&\quad\vrule#\quad&
\hfil#&\quad\vrule#\cr
\noalign{\hrule} 
& &\multispan5\hfil 3.2 Solutions for $m_c=2$  \ \ 
\hfil & &\hfil &
 \cr
\noalign{\hrule} 
& & \omit\hidewidth  $ E_8\otimes A_2 $ \hidewidth& & \omit\hidewidth
$\beta $ \hidewidth& &
\omit\hidewidth  $\beta^2$ \hidewidth 
& & \omit\hidewidth charge \hidewidth & \cr 
\noalign{\hrule}
& &  $\lambda_7\otimes$(0,1)  &  &$\beta=(1,1,1,2,2,2,2,2,2,1,0,1)$ &
&2 &&$Z^{a i}$ &\cr
\noalign{\hrule} 
& & $\lambda_1\otimes$(0,1)  &  &$\beta=(1,1,1,2,2,3,4,5,6,4,3,2)$ &
&0 &&$Z^{a i_1\ldots i_7}$ &\cr
\noalign{\hrule} 
& & $\lambda_1\otimes$(2,0)  &  &$\beta=(1,0,1,2,2,3,4,5,6,4,3,2)$ &
&2 &&$Z_{(a b)}^{i_1\ldots i_7}$ &\cr
\noalign{\hrule} 
& & $1\otimes$(2,0)  &  &$\beta=(1,0,1,2,4,6,8,10,12,8,4,6)$ &
&0 &&$Z_{(a b)}$ &\cr
\noalign{\hrule} 
& & $1\otimes$(0,1)  &  &$\beta=(1,1,1,2,4,6,8,10,12,8,4,6)$ &
&-2 &&$Z^a $ &\cr
\noalign{\hrule} 
& & $1\otimes$(1,2)  &  &$\beta=(1,0,0,2,4,6,8,10,12,8,4,6)$ &
&2 &&$Z_a{}^{(bc)}$ &\cr
\noalign{\hrule} 
}} 
$$
\par
The first multiplet in the above table is a first rank tensor of SL(3) and
so a vector of the Lorentz algebra SO(1,2). It also transforms as the
$E_8$ representation with highest weight state $\lambda_7$
which is the 3875 dimensional representation. When decomposed into $A_8$
representations the 3875 contains $8+70+(216+8)+(28+36+420)+\ldots $. We
note, from the above table,  that it arises from  an 
$E_8^{++++}$ root  that is related to the root $(1^9,0,0,1)$ for
$Z^{1011}$ of equation (3.4) by the  addition of $(0^3,1^7,0,0)$ to the
latter. As such, we find that this multiplet has a highest weight state
that corresponds to $Z^{311}$. This  is just part of $Z^{ai}$ which are
the first 
$8$ of SL(8) in the decomposition of the 3875 representation.  Carrying
out the construction of the
$E_8$ root string  on $\lambda_7$ we find that the next highest $A_{8}$
state in the multiplet has an $E_8^{++++}$ root of $(1^7,2,3,2,1,2)$. On
the other hand,   the charge $Z^{7891011}$  of equation (3.4) has an 
$E_8^{++++}$ root of $(1^7,2,3,2,1,2)$. The difference between the roots
is just that required to convert  $Z^{7891011}$ to  $Z^{3891011}$.
This is part of the $70$ states of $Z^{ai_1\ldots i_4}$ of the eleven
dimensional theory which we identify with the next states in the 
 $\lambda_7$ representation when decomposed into  into
$A_8$ multiplets.  Proceeding in this way we can find all  of the 3875
representation of 
$E_8$ in terms of $A_8$ representations and identify their  origin in the 
eleven dimensional theory. 
\par
Thus we have found that the $l_1$ representation contains the point
particle and string multiplets of brane charges found in references
[15-18] deduced from  the action of  U-duality transformation on known
brane charges.  However, the $l_1$ representation contains an infinite
number of $E_8$  multiplets and and so we can expect that all brane
charges will be packaged together in this representation.

\par

The solutions for $m_c=3$ to equations (2.2.2), (2.2.3) and (2.2.5) for
the case of $G=E_8$ are listed in table 3.3 given below. 

$$\vbox{ \offinterlineskip\halign{
\strut#&\vrule#\quad& 
\quad\quad\hfil#\hfil&
\quad\vrule#\quad&
\hfil#\hfil&
\quad\vrule#\quad&
\it#\hfil&\quad\vrule#\quad&
\hfil#&\quad\vrule#\cr
\noalign{\hrule} 
& &\multispan4\hfil 3.3   Solutions for $m_c=3$  \ \ 
\hfil & &\hfil &
& \cr
\noalign{\hrule} 
& & \omit\hidewidth $E_8\otimes A_2$ \hidewidth& & \omit\hidewidth
$\beta$\hidewidth& &
\omit\hidewidth $\beta^2$\hidewidth 
& & \omit\hidewidth charge \hidewidth & \cr 
\noalign{\hrule}
& & $\lambda_8\otimes$(1,0)  &  &$\beta=(1,1,2,3,3,3,3,3,3,2,1,1)$ &
&2 &&$Z_a^{}$ &\cr
\noalign{\hrule} 
& & $\lambda_2\otimes$(0,2)  & 
&$\beta=(1,1,1,3,3,3,4,5,6,4,2,3)$ & &2 &&$Z^{a b}{}^{i_1\ldots i_6}$
&\cr
\noalign{\hrule} 
& & $2\lambda_1\otimes$(1,0)  & 
&$\beta=(1,1,2,3,2,3,4,5,6,4,2,3)$ & &2 &&$Z_a^{i_1\ldots i_7,j_1\ldots
j_7}$ &\cr
\noalign{\hrule} 
& & $\lambda_2\otimes$(1,0)  &  &$\beta=(1,1,2,3,3,3,3,3,3,2,1,1)$ &
&0 &&$Z_a^{ i_1\ldots i_6}$ &\cr
\noalign{\hrule} 
& & $\lambda_7\otimes$(2,1)  &  &$\beta=(1,0,1,3,4,5,6,7,8,5,2,4)$ &
&2 &&$Z_{(a b)}{}^{ci}$ &\cr
\noalign{\hrule} 
& & $\lambda_7\otimes$(0,2)  &  &$\beta=(1,1,1,3,4,5,6,7,8,5,2,4)$ &
&0 &&$Z^{(a b)i}$ &\cr
\noalign{\hrule} 
& & $\lambda_7\otimes$(1,0)  &  &$\beta=(1,1,2,3,4,5,6,7,8,5,2,4)$ &
&-2 &&$Z_a^{ i}$ &\cr
\noalign{\hrule} 
& & $1\otimes$(1,0)  &  &$\beta=(1,1,2,3,6,9,12,15,18,12,6,9)$ &
&-6 &&$Z_a$ &\cr
\noalign{\hrule} 
& & $1\otimes$(0,2)  &  &$\beta=(1,1,1,3,6,9,12,15,18,12,6,9)$ &
&-4 &&$Z^{(ab)}$ &\cr
\noalign{\hrule} 
& & $1\otimes$(2,1)  &  &$\beta=(1,0,1,3,6,9,12,15,18,12,6,9)$ &
&-2 &&$Z^{a}{}_{(bc)}$ &\cr
\noalign{\hrule} 
& & $1\otimes$(1,3)  &  &$\beta=(1,0,0,3,6,9,12,15,18,12,6,9)$ &
&2 &&$Z^{(abc)}{}_{d}$ &\cr
\noalign{\hrule} 
& & $\lambda_1\otimes$(1,0)  &  &$\beta=(1,1,2,3,4,6,8,10,12,8,4,6)$ &
&-4 &&$Z_a^{i_1\ldots i_7}$ &\cr
\noalign{\hrule} 
& & $\lambda_1\otimes$(0,2)  &  &$\beta=(1,1,1,3,4,6,8,10,12,8,4,6)$ &
&-2 &&$Z^{(ab),i_1\ldots i_7}$ &\cr
\noalign{\hrule} 
& & $\lambda_1\otimes$(2,1)  &  &$\beta=(1,0,1,3,4,6,8,10,12,8,4,6)$ &
&0 &&$Z_{(ab)}^{c,i_1\ldots i_7}$ &\cr
\noalign{\hrule} 
}} 
$$
\par
We can use the equations in this paper to find the reduction to three
dimensions of the non-linearly realised $E_8^{+++}$ theory itself. 
In this case we just consider the decomposition
 of the adjoint representation of $E_8^{+++}$ into 
$E_8\otimes A_2$ representations and so instead of taking $m_*=1$, as we
did for the $l_1$ representations, we take $m_*=0$. As noted in section
two, there are no solutions to the $A_2$ equation (2.23) for $m_c=0,1$,
but there are two solutions for  $m_c=2$ which are given in table (2.2).
We must combine these with the solutions of the $E_8$ equation (2.2.2)
given in
 table (2.3). We then find that the solutions for $m_c=2$ for the highest
weights of 
$E_8\otimes A_2$ in the decomposition of the adjoint representation of 
 $E_8^{+++}$ are
given in the table (3.4) below. 

$$
\vbox{ \offinterlineskip\halign{
\strut#&\vrule#\quad& 
\quad\quad\hfil#\hfil&
\quad\vrule#\quad&
\hfil#\hfil&
\quad\vrule#\quad&
\it#\hfil&\quad\vrule#\quad&
\hfil#&\quad\vrule#\cr
\noalign{\hrule} 
& &\multispan5\hfil 3.4 Solutions for $m_c=2$  \ \ 
\hfil & &\hfil &
 \cr
\noalign{\hrule} 
& & \omit\hidewidth  $ E_8\otimes A_2 $ \hidewidth& & \omit\hidewidth
$\beta $ \hidewidth& &
\omit\hidewidth  $\beta^2$ \hidewidth 
& & \omit\hidewidth charge \hidewidth & \cr 
\noalign{\hrule}
& &  $\lambda_1\otimes$(0,2)  &  &$\beta=(0,0,0,2,2,3,4,5,6,4,2,3)$ &
&2 &&$F^{(ab) i_1\ldots i_7}$ &\cr
\noalign{\hrule} 
& & $1\otimes$(0,2)  &  &$\beta=(0,0,0,2,4,6,8,10,12,8,4,6)$ &
&0 &&$F^{(ab)}$ &\cr
\noalign{\hrule} 
& & $\lambda_1\otimes$(1,0)  &  &$\beta=(0,0,1,2,2,3,4,5,6,4,2,3)$ &
&0 &&$F_{a }^{i_1\ldots i_7}$ &\cr
\noalign{\hrule} 
& & $1\otimes$(1,0)  &  &$\beta=(0,0,1,2,4,6,8,10,12,8,4,6)$ &
&-2 &&$F_{a }$ &\cr
\noalign{\hrule} 
& & $\lambda_7\otimes$(1,0)  &  &$\beta=(0,0,1,2,2,2,2,2,2,1,0,1)$ &
&2 &&$F_a{}^i $ &\cr
\noalign{\hrule} 
}} 
$$
\par
Examining their $E_8^{++++}$ roots we find that the last entry in the
table arises from the eleven dimensional field $A_{2311}$ and is the
first component of the 3875 dimensional representation of $E_8$. 
The first and third entries, at level $n_8=3$, both arise from the eleven
dimensional field 
$h^{a_1\ldots a_8,b}$  and in particular from the states 
$h^{35\ldots 11,3}$ and $h^{35\ldots 11,2}$ respectively. They are in the
adjoint representation of $E_8$ whose first component is an $8$ of SL(8)
and these are given by $h^{i_1\ldots i_7 (a,b)}$ and $h^{i_1\ldots i_7
[a,b]}$ respectively. The first field is a  symmetric second rank tensor 
under SL(3), but this becomes a reducible representation under the
Lorentz algebra SO(1,2). The trace part is just  the usual adjoint
representation of $E_8$ scalar fields  one finds in the
dimensional reduction to three dimensions of the eleven dimensional
supergravity theory. A more detailed analysis of these fields, their
dynamics  and their relationship to the eleven dimensional theory will be
given elsewhere.

%%%%%%%%%%%%%%%%%%%%%%%%%%%%%%%%%%%%%%%%%%%%%%%%%%%%%%%%%%%%%
\medskip
{{\bf 4 Brane charges in nine dimensions and the relationships between
the IIA and IIB theories}}
\medskip
Another interesting dimension in which to compute the brane charges is
nine space-time dimensions as this is the highest dimension in which the
IIA and IIB  supergravity theories dimensionally reduced on a torus
coincide. We will first study this reduction from the point of view of
the eleven dimensional theory, or equivalently, the IIA perspective.  The
decomposition  relevant to the dimensional reduction to nine space-time
dimensions on a torus corresponds to the 
 deletion of the  9th node along the gravity line of the eleven
dimensional theory in the equations  of section two applied
to the case of $E_8^{+++}$. 
The two  algebras arising from the resulting Dynkin diagram are $A_9$ and
$A_1$. The last node of $A_9$ is the exceptional node and hence only the
first eight dots of the $A_9$ are on the gravity line of the nine
dimensional theory and they correspond to the 
$A_{8}$ associated with space-time.  
\par
Adopting the notation of section two the roots and weights of $A_9$ are
defined to be $\beta_i$ and $\mu_i$ for $i=1,2,\ldots ,9$ respectively and
for  the $A_1$ are $\alpha$ and $\lambda$. For this case,
$\alpha_c=x-\nu$ where 
$\nu=\mu_8+\lambda$. One finds that $x^2=-{1\over 10}$.  Equation
(2.2.13)  become
$$
\sum_i p_i (A^{A_9})^{-1}_{ij}-m_c
(A^{A_9})^{-1}_{8j}-m_*(A^{A_9})^{-1}_{1j}=-m_j
\eqno(4.1)$$
where $ p_i, m_j$ are positive integers and the fixed levels $m_c$ and $
m_*$ are also positive integers. Equation (2.2.14) takes the simple
form
$$p=-2m+m_c
\eqno(4.2)$$
where  $p,m$ are positive integers and the level $m_c$ is also a fixed
positive integer. The latter equation is effectively already solved as
one takes all values of
$p$ such that the right-hand side is positive. 
\par
The length squared of the $E^{++++}$ root is given by equation (2.1.11) 
becomes 
$$\beta^2={3\over 2}m_*^2-{(m_c+2m_*)^2\over 10}+{p^2\over 2}
+\sum_{i,j=1}^9 p_i((A^{A_9})^{-1})_{ij}p_j=2,0,-2,\ldots
\eqno(4.3)$$
\par
We are interested in the $l_1$ representation of $E_{11}$ and so we take 
$m_*=1$. The brane charges are classified in terms of $A_9\otimes
A_1$, however,  only the $A_8$ sub-algebra of the $A_9$ is on the gravity
line. We denote the generators of $A_9$ by 
$\hat K^{ a}{}_{ b},\ a,b=1,\ldots , 10$  where 
$a=1,\ldots, 9$, are the indices corresponding  to those of the nine
dimensional space-time, and the final possible index value $ {10}$ 
refers to the exceptional node in the
$E_8^{+++}$ which belongs to   the
$A_9$ Dynkin sub-diagram and does not have any connection with the
space-time index $10$.
The generators of $A_9$  are given, up to a factor,  by 
$$\hat K^{ a}{}_{ b}= K^{ a}{}_{ b},\ {\rm for }\ 
a,b=1,\ldots,9\ {\rm and }\  \hat  K^{ a}{}_{  {10}}=R^{a1011},\ 
\hat K^{  {10}}{}_{ a}=R_{a1011},\ {\rm for }\ 
a=1,\ldots,9 
\eqno(4.4)$$
where $ K^{ a}{}_{ b}$, $R^{a1011}$ and $R_{a1011}$ are the usual
generators of $E_{8}^{+++}$ of the eleven dimensional theory. 
We place hats on the all objects associated with the representations of 
$A_9\otimes A_1$. 
As explained in appendix B, the Cartan involution invariant sub-algebra of
$A_{8}$ is essentially the eight dimensional Lorentz algebra. 
 The internal symmetry is just
$A_1$. 
\par
For $m_c=0$ we find only one solution of equation (4.1), namely 
$p_1=1$, all other Dynkin indices vanishing, and only one solution to
equation (4.2), namely $p=0$. This is an $A_9$ nine rank tensor, or a
tensor with one lower index, but an
$A_1$ singlet, 
$\hat P_{ a}$. The corresponding
$E_{8}^{++++}$ root is 
$\beta=(1,0^{11})$ which has length squared two. As such, we 
recognise  $\hat P_{ a}=P_a, \ a=1,\ldots ,9$ and  
$$\hat P_{10}=-[ \hat K^{ a}{}_{  {10}}, P_{\tilde a}]=-[R^{a1011},
P_a]= -2 Z^{10 11}
\eqno(4.5)$$
we conclude that $P_{ {10}}=2Z^{10 11}$ where $Z^{10 11}$ is the second
member of the $l_1$ multiplet of the $E_{11}$ viewed from the eleven
dimensional perspective. It is just the two rank central charge of the
eleven dimensional supersymmetry algebra. Here we have used the
commutators of reference [22]. As such, for $m_c=0$ the $l_1$
representation decomposed to
$A_9\otimes A_1$  contains the object $\hat P_a$ which consists of the
eleven dimensional generators 
$$
P_a,\ a=1,\ldots,9,\    \  {\rm and}\  Z^{10 11}
\eqno(4.6)$$. 
\par
For $m_c=1$ we only find the   solution of equation (4.1)
namely 
$p_9=1$ all other Dynkin indices vanishing and only one solution to
equation (4.2), namely $p=1$. This is a $A_9$ one rank tensor and an
$A_1$ doublet, 
$\hat Z^{ a i}, \ a=1,\ldots ,10,\ i=1,2$. The corresponding
$E_{8}^{++++}$ root is 
$\beta=(1^{10},0,0)$ which has length squared two and corresponds to  
the highest weight state $\hat Z^{10\  2}$. Using similar arguments
to those deployed for the previous solution we find that $\hat Z^{a i}, \
a=1,\ldots ,10,\ i=1,2$ contains, up to factors, 
$$ Z^{{10}\  2}= P_{10},\ Z^{{10}\  1}= P_{11},\ \hat Z^{ a \ 2}=Z^{a\
11},
\ \hat Z^{ a \ 1}=Z^{a\ 10},  a=1,\ldots,9, i=1,2
\eqno(4.7)$$
\par
For $m_c=2$, we find the   solution of equation (4.1) which consists of  
$p_7=1$, all other Dynkin indices vanishing, together with the solution   
$p=0$ to equation (4.2). This representation is an $A_9$ third  rank
tensor and an
$A_1$ singlet, 
$\hat Z^{ a_1  a_2  a_3 },\ \  a_1,a_2,a_3=1,\ldots ,10$. The
corresponding
$E_{8}^{++++}$ root is 
$\beta=(1^{8},2,2,1,1)$ which has length squared two and corresponds to
the highest weight field $\hat Z^{91011}$. One finds, up to factors  that 
$\hat Z^{ a_1  a_2  a_3 },\ \  a_1,a_2,a_3=1,\ldots ,10$ consists of 
$$\hat Z^{
a_1  a_2  {10} }=Z^{a_1a_2},\ 
\hat Z^{ a_1  a_2  a_3 }=Z^{a_1a_2a_3 10 11},\ \  a_1,a_2,a_3=1,\ldots ,9
\eqno(4.8)$$
\par
In fact, there are  also
another solutions with
$m_c=2$ to equations (4.1) and (4.2) but these does not satisfy equation
(4.3). We have also is discarded other solutions to the above 
equations which do not have   roots in 
$E_{8}^{++++}$ which  do not have a connected support on its Dynkin
diagram. As discussed above, these do not correspond to actual roots of 
$E_{8}^{++++}$. 
\par
The above calculation computed the brane charges in nine dimensions
of the eleven dimensional, or equivalently, the IIA theory in ten
dimensions. However, we could also have considered the
dimensional reduction to nine dimensions starting from the IIB
non-linearly realised theory and reducing on a circle.  The difference
between the
$E_8^{+++}$ formulations that lead to the IIA and IIB theories is that
the $A_9$  subalgebras  associated with gravity 
are identified differently [20,27,32,28]. For the IIA case, the $A_9$
gravity line contains the nodes labeled 1 to 9 along the horizontal line
of the $E_8^{+++}$ Dynkin diagram starting from the very extended node.  
While for the IIB theory, the $A_9$
gravity line of the $E_8^{+++}$ Dynkin diagram contains the nodes labeled
1 to 8 which are  along the horizontal line of the Dynkin diagram starting
form the very extended  node and the exceptional node labeled 11. To
find the brane charges in the IIB theory in ten dimensions requires
calculating the content of the
$l_1$ representation in terms of the $A_9$ representation associated with
the gravity line. This means we must consider the $E_8^{++++}$ algebra
with level one on the extra node, labeled * and then delete node ten. The
brane charges are then classified in terms of the remaining $D_{10}$
algebra. However, it is  more convenient to consider a further
decomposition by deleting node 9, whereupon the brane charges are
classified by 
$A_9\otimes A_1$ and labeled by the two levels $m_9, m_{10}$.  However,
this is in effect what we have just done in the calculation above,  we
deleted node 9  and the remaining algebra was $A_9\otimes A_1$. The
$A_9$ algebra is just that of the gravity line and so corresponds to the
ten dimensional space-time in the IIB theory and the $A_1$ algebra has
representations labeled by $p$,  or equivalently $m_{10}$, but the latter
is just  the level
 associated with the node 10. Hence,  we have already carried out the
required decomposition of the $l_1$ in terms of $A_9$ with respect to the
levels $(m_9,m_{10})$. The result was 
$$\hat P_a,\ \hat Z^{ a i}, \ \hat Z^{ a_1  a_2  a_3 },\ldots 
a=1,\ldots ,10,\ i=1,2
\eqno(4.9)$$
We recognise that the  $l_1$ representation of the IIB theory contains at
low levels   the space-time translations and the first two
central charges that occur in the supersymmetry algebra of  the IIB
theory,  Indeed, the hated notation was designed to be suitable for this
interpretation of the calculation. 
\par
As explained in reference [28], the is a one to one relation between the
three $E_8^{+++}$ non-linear realisations associated with the eleven
dimensional theory, the IIA theory and IIB theory which arises from
their common $E_8^{+++}$ origin. Indeed, the above identifications of the
$l_1$ representations in the three theories given in equations (4.6-4.9),
extend those given in this work. These identifications do not
require any compactification of the three theories, but 
they also hold  if the theories are compactified.  In particular, we note
that the the tenth component of the momentum of the IIB theory  $\hat
P^{10}$ is identified with the  component $Z^{10 11}$ of the membrane
charge, or equivalently, the charge of the string in the IIA theory. 
We recall that  the momenta in the compactified dimensions are the charges
associated with the Kaluza-Klein modes and that the topological
charges associated with the winding modes of the string, or membrane,
are the  central charges  in the compactified dimensions that
occur in the supersymmetry algebra.  
In the nine dimensional theory, this means that the winding modes of the 
IIA string on the circle on which  the IIA theory is reduced have a
charge which is the 
$Z^{1011}$ and so this  must be  identified with the Kaluza-Klein
modes of the IIB theory whose charge is $\hat P_{10}$. Similarly, we also
learn form equations (4.7) and (4.9) that the winding modes of the two
stings in the IIB theory, whose charges are $\hat Z^{10 i}$, must be 
identified with the Kaluza-Klein modes of the IIA theory reduced from
eleven dimensions, whose charges are $P_{10}$ and $P_{11}$. 
\par
Thus we recover the results of [39] which studied how the IIA and IIB
supersymmetry algebras in ten dimensions lead to the unique nine
dimensional supersymmetry algebra. As a result of  these different
origins, these authors were able to conclude that the Kaluza-Klein modes
of the IIA string and those of the IIB string belonged to different
supersymmetry multiplets and that T duality mapped the Kaluza-Klein 
modes of the IIA string into the string winding modes of the IIB string
and that the winding modes of the IIA string were mapped to the 
Kaluza-Klein modes of the IIB theory. It is encouraging
that these T- duality rules  can be read off from their common
$E_8^{+++}$ origin in a simple way.   Particularly, given that  these
results were derived in reference [39] using the string properties and
supersymmetry algebra of these theories and these features have so far yet
to be identified in the $E_8^{+++}$ approach. 
\par
In reference [22] it was proposed to consider the non-linear
realisation of the semi-direct product representation of $E_8^{+++}$ and
its $l_1$ representation. In this approach one has fields which are in a
one to one correspondence with the generators of   the
Borel sub-algebra of
$E_8^{+++}$, but also generalised coordinates which are in a one to one
correspondence with the content of the 
$l_1$ representation.  The fields then depend on
these generalised coordinates. For the eleven dimensional theory, these
generalised coordinates are 
$x^a, z_{ab}, z_{a_1\ldots a_5},\ldots $. As explained in [28], when
making the correspondence between the eleven dimensional, IIA and IIB
theories one must not only swop the fields, but also exchange 
the  generalised coordinates as explained above  at
low levels. 
\par
If we dimensionally reduce both the IIA and IIB non-linearly
realised theories on a circle and keep all their dependence on the
generalised coordinates, that is not only  keep just the   massless modes
for example, we will find two theories in nine dimensions  both of which
have their original number of coordinates. In particular, the IIA theory 
will depend on the generalised coordinates
$x^a, \ a=1,\ldots,  9$, $x^{10}$, $x^{11}$ , $z_{1011}$ and an infinite
number of other coordinates. While, the IIB theory in nine dimensions will
have the generalised coordinates  $x^a, \ a=1,\ldots,  9$, $x^{10}$,
$z_{10\ i}$ and an infinite number of other  coordinates. 
Indeed, if we just
consider the nine dimensional theory without worrying about keeping track
of its  higher dimensional origin then we must decompose the $E_8^{++++}$
algebra  with respect to the algebra that remains by deleting the nodes
labeled  eight, nine, ten and eleven with the corresponding level
$(m_{8},m_{9}, m_{10}, m_{11})$. We note that we delete the same node if
we derive the theory from the IIA and IIB perspective and so the resulting
theory derived from either path will be identical. 
\par
We have arrived at a theory which has some elements in common with that
considered in references [40,19]. These authors  proposed to construct
 what was called a BPS extended nine dimensional theory consisting
of nine dimensional supergravity coupled to the two towers of
Kaluza-Klein multiplets as well a tower of states corresponding to
the Kaluza-klein modes of the IIB string, or equivalently, the winding
modes of the IIA string. 
 Given the relations between these towers explained above, this 
nine dimensional theory is automatically T duality invariant.  Given the
charges associated with the towers it was proposed [19] that such a
theory would arise by encoding three extra coordinates $x^{10}$, $x^{11}$
and $z_{10 11}$.  However, this is just a restriction of the theory that
would result from the dimensional  reduction of the non-linear
realisation of
$E_8^{+++}$.  However, this latter theory, unlike the BPS extended
theory,   would be  ten dimensional Lorentz invariant. 
One may turn all this around and interpret this as evidence for the 
method of encoding space-time advocated in [22].  
\par
We close this section with a comment of the role of $D_{10}$. 
It is obvious that  the $E_8^{+++}$ non-linearly realised theory also
possess a
$D_{10}$  symmetry as this algebra is obtained by deleting the node
labeled 11 of the $E_8^{+++}$ Dynkin diagram. This node is the last node
in the gravity line associated with the eleven dimensional theory. As
such, examining the $D_{10}$ decomposition of the eleven dimensional theory
by deleting this node destroys the manifest
$A_{10}$, or SL(11), symmetry of the theory. Indeed, 
deleting this node  leads to a residual gravity line with nine
dots, or $A_9$, which is just the gravity line of the  IIA theory.
In fact, it is by deleting this node and examining the $E_8^{+++}$ content
in terms of the remaining algebra  that gives   the fields or generators 
IIA theory. Hence,   it is natural to  formulate the
IIA theory in terms   its
$D_{10}$ symmetry with the
$ A_9$ gravity line being part of this symmetry.  Let us  consider the
$l_1$ representation from this perspective;  its first components,  
$P_a$ are the usual space-time translations and must be part of the
fundamental representation of
$D_{10}$ associated with its  first node, that is the one  labeled one
in the
$E_8^{+++}$ Dynkin diagram. However, this representation also contains
 a state
$ Q^a$ which from the $A_9$ viewpoint has level one with respect to
the one node not on the $A_9$ line of  the $D_{10}$ Dynkin diagram. In
fact, it has an $E_8^{++++}$ root of $(1^9,0,0,1)$.  This is just the 
central charge $ Z^a$ in the IIA supersymmetry algebra which arises
from the eleven dimensional central charge
$Z^{a11}$. 
\par
Deleting the node labeled 10 in the 
$E_8^{+++}$ Dynkin diagram to find $D_{10}$ also preserves  a different 
$ A_9$, denoted $\hat A_9$, which consists of the nodes labeled one
to eight as well as node eleven. As explained above,  this is just the
$\hat A_9$  gravity line of the IIB theory and so this theory can also be
formulated in terms of
$D_{10}$  which includes its $A_9$  gravity line. However, to deduce the
$\hat  A_9$ content of the theory we delete a different
node of the
$E_8^{+++}$ Dynkin diagram as we did  in the IIA case, namely the node
labeled nine  in the $E_8^{+++}$ Dynkin diagram which is connected to the
node we already delete to expose the $D_{10}$ symmetry. Now the
$l_1$ representation contains the  usual space-time translations  $\hat
P_a$ of the IIB theory as well as  a state 
$\hat Q^a$.  However, in contrast to the IIA case, the highest weight
state of $\hat Q^a$ has 
$E_8^{++++}$ root $(1^{10},0,0)$ and it is identified with the central
charge $\hat Z^{a 2}$ in the ten dimensional supersymmetry algebra.
%%%%%%%%%%%%%%%%%%%%%%%%%%%%%%%%%%%%%%%%%%%%%%%%%%%%%%%%%%%%%%%%%%%%%%%%%%
\medskip
{{\bf 5 Discussion and conclusion }}
\medskip
In this paper we have derived equations which can be used to
determine  the decomposition of
$G^{+++}$ and its fundamental representation $l_1$, associated with the
very extended node, into  the sub-algebra whose Dynkin diagram is the
one obtained from $G^{+++}$ by deleting   a node on the gravity
line. These are the sub-algebras that arise  when the non-linearly 
realised theory
is dimensionally reduced. Following [23]
we used the extended Dynkin diagram obtained by adding a further node to
the very extended node by a single line to derive the content of the
$l_1$ representation. Indeed, the level one, with respect to the
new node, states in the adjoint representation of the extended algebra
form the $l_1$ representation of $G^{+++}$ . In fact, this technique can
be used to discuss any  representation of
$G^{+++}$. If the representation has   Dynkin indices
$p_j$ we  just add a new node with $p_j$ lines to the jth node of
$G^{+++}$ and consider the level one generators of this new algebra.
Indeed, one may apply this method to  any Kac-Moody algebra and not just
very extended algebras. 
\par
We used the results of section two in section three to compute the
contents of the $l_1$ representation in terms of $E_8\otimes A_2$. This
is the same as computing the $E_8$ multiplets of brane charges that occur
when the non-linearly realised theory is reduced on an eight torus to
three dimensions. At the lowest level, the point particle and string
multiplets of charges we find are agreement with the previous results of
references [15-18]  which were derived by starting from some of the known
charges and computing  the remainder assuming that  the U-duality
transformations hold. The form of the U-duality transformations are just
the T-duality transformations of string and the assumed $SL(2,{\bf Z})$
transformation of the IIB theory.  However, as the authors of references
[15-19] pointed out one finds in these multiplets many exotic charges
which do not seem to arise from  M theory as it has been previously
discussed. However, from the view point of the eleven dimensional
non-linearly realised 
$E_8^{+++}$ theory their origin is clear. As observed in reference [23], 
there is a correspondence between the brane charges in the $l_1$
representation and the fields that appear in the non-linear realisation 
and hence for a given charge we know the field to which it couples. 
The mysterious  charges can then be seen to couple to fields that are
beyond those that occur in the supergravity approximation. 
\par
In the $l_1$ representation there are an infinite number of charges  
and so an infinite number of $E_8$ representations, however, it is 
encouraging to see how the different types of branes, i.e. point
particles, strings, etc,  all package up into this representation.  the 
sympathetic reader  can interpret these results  as further evidence  
for the $E_{11}$ symmetry underlying M theory. 
\par
In section four, we performed a similar calculation for the nine
dimensional theory and in particular traced how the brane charges arose
from their different IIA and IIB  origins. At low levels we find
agreement with the correspondences found  in references [19,39,40] using
string theory and the supersymmetry algebra. Furthermore, adopting the
approach to space-time  advocated in reference [22] and making a  
considerable truncation we find a nine dimensional theory that makes
contact with  the type of BPS extended theory envisaged in [40,19]. 
\medskip
{{\bf A Weights and Inverse Cartan Matrix of $E_n$}}
\medskip
The reader is invited to draw the Dynkin diagram of $E_n$. 
 We draw    $n-1$ dots connected by a  horizontal line 
 and then placing another dot (the exceptional node) above
the third node from the right and connecting it with a single line to
that node. We label the nodes in the horizontal line by $1,2,\ldots, n-1$ 
from left to right and the node above the line by $n$. 
Following closely the techniques of  reference [30], we
use the decomposition of
$E_n$ to $A_{n-1}$ given by deleting the exceptional node $n$. Let 
$\alpha_i$ and $\lambda_i$ for $i=1,\ldots , n-1$ be the simple roots and
fundamental weights respectively of $A_{n-1}$. 
The roots of
$E_n$  can be  written as [30]
$$\alpha_i, \ i=1,\ldots , n-1, \ \alpha_n=x-\lambda_{n-3}
\eqno(A.1)$$
where $x$ is orthogonal to the roots of $A_{n-1}$ and 
$x^2={(9-n)\over n}$ in order that $\alpha_n^2=2$. 
The fundamental weights of $E_n$ are
given by 
$$l_i=\cases{\lambda_i+{3i\over (9-n)}x,\quad i=1,\ldots,n-3\cr
\lambda_i+{(n-3)(n-i)\over (9-n)}x,\quad i=n-3,\ldots,n-1\cr}
\eqno(A.2)$$
and 
$$
l_n={n\over (9-n)}x
\eqno(A.3)$$
In deriving this result we used the scalar products of the fundamental
weights of $A_{n-1}$
$$
\lambda_i\cdot \lambda_j=(A^{A_{n-1}})^{-1}_{ij}= {i(n-j)\over n},\ 
{\rm for }\ \  i\le j
\eqno(A.4)$$
\par
The inverse Cartan matrix of $E_n$ is given by 
$$((A^{E_n})^{-1})_{ab}=l_a.l_b . 
\eqno(A.5)$$
Using equation (A.4), we find the following algebraic formulae for the
inverse Cartan matrix of $E_{n}$
$$((A^{E_n})^{-1})_{ij}=\cases {{i(9-n+j)\over (9-n)},\quad
i,j=1,\ldots,n-3, i\le j\cr
{(n-j)((n-3)^2-i(n-5))\over (9-n)}),\quad i,j=n-3,\ldots,n-1, i\le j
\cr
2{i(n-j)\over (9-n)}, \quad i=1,\ldots,n-3, j=n-3,\ldots,n-1, 
\cr}
\eqno(A.6)$$
and 
$$
((A^{E_n})^{-1})_{in}=\cases {{3i\over (9-n)},\quad  i=1,\ldots,n-3\cr 
{(n-3)(n-i)\over (9-n)},\quad  i=n-3,\ldots,n-1\cr}
\eqno(A.7)$$
and 
$$
((A^{E_n})^{-1})_{nn}= {n\over (9-n)}
\eqno(A.8)$$
\par
It is easy to check that for $n=8,10$ the inverse Cartan matrix has 
integer valued entries,  if $n\le 8$ they are  positive and for
$n=10$ they are  negative.

\medskip
{\bf B Cartan involution invariant sub-algebra of $A_n$ and $G^+$}
\medskip
Given a Kac-Moody algebra, the Cartan involution is defined to act on the
Chevalley generators $E_a$, $F_a$ and $H_a$ as 
$$E_a\to -F_a,\ F_a\to -E_a, H_a\to -H_a
\eqno(B.1)$$
As such, the sub-algebra invariant under the Cartan involution is 
generated by 
$$ E_a-F_a
\eqno(B.2)$$
\par
The generators of $A_n$ are $K^a{}_b, \ a,b=1,2,\ldots ,n+1$ and obey the 
relations 
$$[K^a{}_b,K^c{}_d]=\delta _b^c K^a{}_d - \delta _d^a K^c{}_b .
\eqno(B.3)$$
The Chevalley generators are 
$$E_a=K^a{}_{a+1},\ F_a=K^{a+1}{}_{a},\ H_a=K^a{}_{a}-K^{a+1}{}_{a+1}
\eqno(B.4)$$
The reader may readily verify that they do satisfy the defining relations
of the Kac-Moody algebra corresponding to the Cartan matrix of $A_n$. 
\par
The Cartan involution defined above induces an action on all the
generators of $A_n$ as follows; 
$$K^{a}{}_{b}\to -K^{b}{}_{a}
\eqno(B.5)$$
and as such the Cartan involution invariant sub-algebra has the
generators 
$$J_{ab}=K^{a}{}_{b} -K^{b}{}_{a}. 
\eqno(B.6)$$
It is straightforward to verify using equation (B3) that 
$J_{ab}$ obey the commutation relations of $SO(n+1)$. As such, we recover
the  known fact that the Cartan involution invariant sub-algebra of
$SL(n+1)$ is $SO(n+1)$. 
\par
A generator in the vector representation of $A_n$, i.e. with 
non-vanishing Dynkin index
$p_{n}=1$,  transforms under $A_n$  as
$$
[K^a{}_b, R^{c}]= \delta _b^{c}R^{a}-{\delta_b^a\over n+1} R^c,   
\eqno(B.7)$$
with $K^a{}_b$ acting in a similar way for more complicated tensors. 
It is easy to verify that 
$$[J_{ab}, R^{c}]= \delta _b^{c}R^{a}- \delta _a^{c}R^{b}
\eqno(B.8)$$
with a similar action on more complicated tensors. 
Hence, a tensor under SL(n+1) transforms under its  
Cartan involution invariant sub-algebra,  SO(n+1) as the indices
suggest.  However, as  $\delta_{ab}$ is an $SO(n+1)$ invariant tensor, the
representation of $SL(n+1)$ is not always an irreducible representation
of $SO(n+1)$ . For example, the tensor
$T^{(ab)}$ of  $SL(n+1)$  with non-vanishing Dynkin index $p_{n}=2$
transforms irreducibly under $SO(n+1)$ i.e. as singlet  and a symmetric
traceless tensor, 
$T^{(ab)}-{\delta^{ab}\over n+1} T^{c}{}_c$. 
\par
By introducing a suitable number of minus signs into  the Cartan involution
of equation (B.1) and repeating the above steps one finds that the
invariant sub-algebra is $SO(p,n+1-p)$. Rather than continually record
these minus signs, it is more efficient to just remember that one should 
put them in,   but to stick to the signs of equation (C.1) and simply
then restore the final result to what it should be by  simply carry
out a Wick rotation on the final result. 
\par
The above discussion plays an important role in this paper as the 
space-time Lorentz algebra arises in the non-linear realisation just as
the above suggests,  namely as the Cartan involution invariant
sub-algebra of the $A_D$ associated with the gravity line. As such, the
Lorentz algebra in the dimensionally reduced theory is just the 
Cartan involution invariant
sub-algebra of the $A_D$ associated with the gravity line which remains
to the left of the node that is deleted. The representations of the
Lorentz group are then read off from those of $SL(n+1)$ as 
described above.  
\par
In the remainder of this appendix we give the Cartan involution
invariant sub-algebra of
$G^+$.  This is work carried out with Matthias Gaberdiel and a more
detail account will be given elsewhere [41]. Let us consider any finite
dimensional semi-simple Lie algebra $G$ with generators $E_\alpha$ and
$H_a$ where
$\alpha$ is any root. Under the Cartan involution of equation (B.1) 
these transform as $E_\alpha \to -E_{-\alpha}$ and $H_a\to -H_a$. The
roots of the affine algebra $G^+$ can be written in the form
$(\alpha,0,n)$ and $(0,0,n)$ and under the Cartan involution they
transform as  to $(-\alpha,0,-n)$ and $(0,0,-n)$ respectively. The
corresponding generators are $E_{\alpha, n}$ and $H_{a,n}$ and, together
with the central generator $k$,  they transform as 
$$
E_{\alpha, n}\to -E_{-\alpha, -n},\ 
H_{a,n}\to -H_{a,-n}\ {\rm and } \ k\to -k
\eqno(B.9)$$
Taking the loop approach to the affine algebra we may write the generators
in the form 
$$
E_\alpha (x)=\sum _n E_{\alpha, n} x^{-n},\ 
H_{a} (x)=\sum _n  H_{a,n} x^{-n}
\eqno(B.10)$$
where $x=e^{i\theta}$ for $-\pi\le\theta \le \pi$. Under the Cartan
involution 
$$
E_\alpha (x)\to -E_{-\alpha} (x^{-1}),\ 
H_{a} (x)\to -H_{a} (x^{-1})
\eqno(B.11)$$
Hence, the Cartan involution invariant sub-algebra contains   the
combinations 
$$
K_\alpha (x)=E_\alpha (x)-E_{-\alpha} (x^{-1}),\ {\rm and }\ 
L_a (x)=H_{a} (x)-H_{a} (x^{-1})
\eqno(B.12)$$
We note that the central generator has been eliminated. 
It is more useful
to work with  combinations that have a definite symmetry under $\theta\to
-\theta$ and so we consider the 
 Cartan involution invariant sub-algebra to contain 
$$
Q_\alpha= K_\alpha (x)-K_\alpha (x^{-1}),\ 
P_\alpha= K_\alpha (x)+K_\alpha (x^{-1}),\ L_a(x)
\eqno(B.13)$$
\par
We can construct the 
Cartan involution invariant sub-algebra of $G^+$ directly from $G$ by
considering an interval algebra rather than the loop algebra which leads
to  $G^+$. We divide the generators of $G$ into those that are
eigenstates of the Cartan involution, namely 
$P_\alpha=E_\alpha -E_{-\alpha}$, $Q_\alpha=E_\alpha +E_{-\alpha}$ and 
$H_a$. We then consider the mapping from the interval $[0,\pi]$ into the
group $G$ to definite the generators 
$P_\alpha(\theta),\ Q_\alpha(\theta),\ H_a(\theta)$ 
which are subject to the boundary conditions 
$${d P_\alpha(\theta)\over d\theta}=0, Q_\alpha(\theta)=0,
H_a(\theta)=0
\eqno(B.14)$$
at $\theta=0,\pi$. Hence,  $P_\alpha(\theta)$ obeys Neumann boundary
conditions while $Q_\alpha(\theta)$ and $H_a(\theta)$ obey
Dirichlet boundary conditions. We can extend the range of the interval to
be from $-\pi$ to $\pi$ by demanding that
$P_\alpha(\theta)=P_\alpha(-\theta),\ 
Q_\alpha(\theta)=-Q_\alpha(-\theta),\ H_a(\theta)=-H_a(-\theta)$. 
\par
In fact,  the 
 Cartan involution invariant sub-algebra 
of $G^+$  is not a Kac-Moody algebra as can be shown by finding the 
invariant bilinear form on the algebra and showing that it is not 
non-degenerate.

%%%%%%%%%%%%%%%%%%%%%%%%%%%%%%%%%%%%%%%%%%%%%%%%%%%%%%%%%%%%%%%%%%%%%%%%
\medskip 
{\bf Acknowledgments}
I wish to thank Axel Kleinschmidt, Boris Pioline and  Bernard de Wit 
 for useful discussions.  I wish also to thank the  organisers of
the Strings in the Pyrenees 2003 workshop at  Benasque, the Erwin
Schršdinger International Institute for Mathematical Physics at Wien and
the Department of Physics at Heraklion for their hospitality.   This
research was supported by a PPARC senior fellowship PPA/Y/S/2002/001/44
and  in part by the PPARC grants  PPA/G/O/2000/00451, 
PPA/G/S4/1998/00613 and the EU Marie Curie, research training network
grant HPRN-CT-2000-00122. 
\medskip
{\bf References}
\medskip
\item{[1]} I.C.G. Campbell and  P. West, {\it N=2 d=10 nonchiral
supergravity
     and its spontaneous compactifications}, Nucl. Phys. {\bf B243} (1984),
     112; M. Huq, M. Namanzie, {\it Kaluza-Klein supergravity in ten
     dimensions}, Class. Quant. Grav. {\bf 2} (1985); F. Giani, M. Pernici,
     {\it N=2 supergravity in ten dimensions}, Phys. Rev. {\bf D30} (1984),
     325
\item{[2]} J. Schwarz and  P. West {\it Symmetries and Transformations of
     chiral N=2, D=10 supergravity}, Phys. Lett. {\bf B126} (1983), 301.
\item{[3]}
     J. Schwarz, {\it Covariant field equations of chiral N=2 D=10
     supergravity}, Nucl. Phys. {\bf B226} (1983), 269; P. Howe and  P.
West,
     {\it The complete N=2, d=10 supergravity}, Nucl. Phys. {\bf B238}
     (1984), 181
\item{[4]} S. Ferrara, J. Scherk and B. Zumino, {\sl Algebraic
properties of extended supersymmetry}, Nucl. Phys. {\bf B 121} (1977)
393; E. Cremmer, J. Scherk and S. Ferrara, {\sl $SU(4)$ invariant
supergravity theory}, Phys. Lett. {\bf B 74} (1978) 61
\item{[5]} E. Cremmer and B. Julia, {\sl The $N=8$ supergravity
theory. I. The Lagrangian}, Phys. Lett. {\bf B 80} (1978) 48
\item{[6]} N. Marcus and J. Schwarz, {\it Three-dimensional
supergravity theories}, Nucl. Phys. {\bf B228} (1983) 301. 
\item{[7]} E. Cremmer, B. Julia and J. Scherk, 
{\it  Supergravity theory in eleven dimensions},  Phys. Lett. 76B
(1978) 409.
\item{[8]} B.\ Julia, {\it Group disintegrations}, in {\it  
Superspace and
Supergravity}, p. 331,  eds. S. W. Hawking  and M.  Ro\v{c}ek,  
Cambridge University
Press (1981).
\item{[9]} M. de Roo, Nucl. Phys. {\bf B255} (1985) 515; 
S. Ferrara, C. Kounnas and M. Porrati, 
Phys.Lett. {\bf B181} (1986), 263; 
J. Maharana, J.H. Schwarz, {\it Noncompact symmetries in
     string theory}, Nucl. Phys. {\bf B390} (1993) 3 {\tt hep-th/9207016},
\item{[10]} S. J. Rey, {\it The confining phase of superstrings and  
axionic strings}, Phys. Rev. {\bf D43} (1991) 526, UCSB-TH-89-23: 
{\it Axionic string instantons and their low energy implications,
UCSB-TH-89-23};  A. Font, L. Ibanez, D. Lust and F. Quevedo, {\it  
Strong-weak coupling duality and nonperturbative effects in string
theory},    Phys.Lett. B249 (1990) 35.  
\item{[11]} C.M. Hull and P.K. Townsend, {\it Unity of  
superstring  dualities},
Nucl.\ Phys.\ {\bf B438} (1995) 109, {\tt arXiv:hep-th/9410167}.
\item{[12]}  K. Kikkawa and M. Yamasaki, {\it Casimir effects  
in
superstring theories}, Phys. Lett. {\bf B149} (1984) 357; N. Sakai and  
I. Senda, {\it
Vacuum energies of string compactified on torus}, Prog. Theor. Phys.  
{\bf 75}
(1986) 692; T. H. Buscher, {\it A symmetry of the string background  
field
equations}, Phys. Lett. {\bf B194} (1987) 59; {\it Path integral  
derivation of
quantum duality in non-linear $\sigma$-models}, Phys. Lett. {\bf B201}  
(1988) 466.
\item{[13]}  M. Rocek and E. Verlinde, {\it Duality, quotients  
and
currents}, Nucl. Phys. {\bf B373} (1992) 630, {\tt  
arXiv:hep-th/9110053}.
\item{[14]} J. Azcarraga, J. Gauntlett, J. Izquierdo and P. Townsend, 
{\it Topological Extensions of the Supersymmetry Algebra for Extended
Objects}, Phys. Rev. Lett. {\bf 63, no 22} (1989) 2443. 
\item{[15]} S. Elitzur, A. Giveon, D. Kutasov and E.  Rabinovici,  {\it
Algebraic aspects of matrix theory on $T^d$ }, {\tt  
arXiv:hep-th/9707217}.
\item{[16]}  N. Obers,  B. Pioline and E.  Rabinovici, {\it M-theory and
U-duality on $T^d$ with gauge backgrounds}, {\tt hep-th/9712084}
\item{[17]} N. Obers and B. Pioline,~ {\it U-duality and  
M-theory, an
algebraic approach}~, {\tt hep-th/9812139}.
\item{[18]} N. Obers and B. Pioline,~ {\it U-duality and  
M-theory}, {\tt arXiv:hep-th/9809039}.
\item{[19]} B. de Wit and H. Nicolai, {\it Hidden symmetries, central
charges and all that}, hep-th/0011239. 
\item{[20]} P. West, {\sl $E_{11}$ and M Theory}, Class. Quant.
Grav. {\bf 18 } (2001) 4443, {\tt hep-th/0104081}
\item{[21]} P.~C. West, {\sl Hidden superconformal symmetry in {M}
    theory },  JHEP {\bf 08} (2000) 007, {\tt hep-th/0005270}
\item{[22]} P. West, {\sl $E_{11}$, SL(32) and Central Charges},
Phys. Lett. {\bf B 575} (2003) 333-342, {\tt hep-th/0307098}
\item{[23]} A. Kleinschmidt and  P. West, {\sl Representations of ${\cal
G}^{+++}$ and the role of space-time}, hep-th/0312247. 
\item{[24]} T. Damour, M. Henneaux and H. Nicolai, {\sl $E_{10}$ and a
``small tension expansion'' of M-theory}, Phys. Rev. Lett. {\bf 89}
(2002) 221601, {\tt hep-th/0207267}
\item{[25]} F. Englert and L. Houart, {\sl ${\cal G}^{+++}$ invariant
formulation of gravity and M-theories: Exact BPS solutions}, {\tt
hep-th/0311255} 
\item{[26]}  F. Englert, L. Houart, A. Taormina and P. West,
{\sl The Symmetry of M-theories}, JHEP 0309 (2003) 020, 
{\tt hep-th/0304206}
\item{[27]} I. Schnakenburg and P. West, {\sl Kac-Moody Symmetries of
IIB supergravity}, Phys. Lett. {\bf B 517} (2001) 137-145, {\tt
hep-th/0107181} 
\item{[28]} P. West, {\it  The  IIA, IIB and eleven dimensional 
theories and their common
$E_{11}$ origin. }, hep-th/0402140. 
\item{[29]}  N. Lambert and  P. West, {\sl Coset symmetries in
dimensionally reduced bosonic string theory}, Nucl. Phys. {\bf B 615}  
(2001) 117, {\tt hep-th/0107209}
\item{[30]}  M. R. Gaberdiel, D. I. Olive and P. West, {\sl A
class of Lorentzian Kac-Moody algebras}, Nucl. Phys. {\bf B 645}
(2002) 403-437, {\tt hep-th/0205068}
\item{[31]} F. Englert, L. Houart and P. West, {\sl
  Intersection rules, dynamics and symmetries}, JHEP 0308 (2003) 025, 
{\tt hep-th/0307024}
\item{[32]} A. Kleinschmidt, I. Schnakenburg and P. West, {\sl
Very-extended Kac-Moody algebras and their interpretation at low
levels}, {\tt hep-th/0309198}
\item{[33]} P. West, {\sl Very-extended $E_8$
and
$A_8$ at low levels}, Class. Quant. Grav. {\bf 20} (2003) 2393, 
hep-th/0307024. 
\item{[34]}  H. Nicolai and T. Fischbacher, {\sl Low Level
representations of $E_{10}$ and $E_{11}$}, Contribution to the
Proceedings of
the Ramanujan International Symposium on Kac-Moody Algebras and
Applications, ISKMAA-2002, Jan. 28--31, Chennai, India, {\tt
hep-th/0301017}
\item{[35]} T. Banks and W. Fischler and L. Motl, {\it  
Dualities versus
singularities}, {\bf JHEP 9901} (1999) 019, {\tt arXiv:hep-th/9811194}.
\item{[36]} V. Kac, {\sl Infinite dimensional algebras}, 3rd edition,
Cambridge University Press (1990)
\item{[37]} A. Shapere, S. Trivendi and F. Wilczek, Mod. Phys. Lett {\bf
A6} (1991) 2677; 
A. Sen, Nucl. Phys. {\bf B404} (1993) 109  {\tt hep-th/9207053} ; Phys.
Lett. {\bf B303} (1993) 22  {\tt hep-th/9209016},  Mod.   Phys. Lett.
{\bf A8} (1993) 2039  {\tt hep-th/9303057}; J.H. Schwarz, A. Sen, {\it
Duality Symmetries of 4d Heterotic 
     Strings}, Phys.Lett. {\bf B312} (1993) 105, {\tt hep-th/9305185};  
 Nucl. Phys. {\bf B411} (1994) 35  {\tt hep-th/9304154}. 
\item{[38]}I. Schnakenburg and P. West, {\it Kac-Moody Symmetries of
Ten-dimensional \ Non-maximal Supergravity Theories}, JHEP {\bf 0405}
(2004) 019 {\tt hep-th/0401196}
\item{[39]}  M. Abou-Zeid, B. de Wit, D. Lust, H. Nicolai, 
{\it Space-Time Supersymmetry, IIA/B Duality and
M-Theory},Phys. Lett. {\it B466} (1999) 144, hep-th/9908169.  
\item{[40]} B. de Wit, {\it M-theory duality and
BPS-extended supergravity}, ,hep-th/0010292. 
\item {[41]} M. Gaberdiel and P. West, to be published.

\end